\documentclass[aps,pra,twocolumn,showpacs,amsmath,amssymb,preprintnumbers,superscriptaddress,nofootinbib,10pt]{revtex4-1}
\usepackage{amsmath,amssymb}
\usepackage{bm}% bold math
\usepackage[latin1]{inputenc}
\usepackage{dcolumn}% Align table columns on decimal point
\usepackage{graphicx}% Include figure files
\usepackage{SIunits}
\usepackage{braket}

\usepackage{hyperref,breakurl} % Uncomment to insert urls.

\begin{document}

\title{Attacks on practical quantum key distribution systems\\(and how to prevent them)}

\author{Nitin Jain}
\email{nitinj1981@gmail.com}
\affiliation{Max Planck Institute for the Science of Light, Erlangen, Germany}
\affiliation{Institute of Optics, Information and Photonics, University of Erlangen-Nuremberg, Germany}
\affiliation{EECS Department, Northwestern University, Evanston, USA}

\author{Birgit Stiller}
\affiliation{Max Planck Institute for the Science of Light, Erlangen, Germany}
\affiliation{Institute of Optics, Information and Photonics, University of Erlangen-Nuremberg, Germany}
\affiliation{Centre for Ultrahigh bandwidth Devices for Optical Systems (CUDOS), School of Physics, University of Sydney, Australia}

\author{Imran Khan}
\affiliation{Max Planck Institute for the Science of Light, Erlangen, Germany}
\affiliation{Institute of Optics, Information and Photonics, University of Erlangen-Nuremberg, Germany}

\author{Dominique Elser}
\affiliation{Max Planck Institute for the Science of Light, Erlangen, Germany}
\affiliation{Institute of Optics, Information and Photonics, University of Erlangen-Nuremberg, Germany}

\author{Christoph Marquardt}
\affiliation{Max Planck Institute for the Science of Light, Erlangen, Germany}
\affiliation{Institute of Optics, Information and Photonics, University of Erlangen-Nuremberg, Germany}

\author{Gerd Leuchs}
\affiliation{Max Planck Institute for the Science of Light, Erlangen, Germany}
\affiliation{Institute of Optics, Information and Photonics, University of Erlangen-Nuremberg, Germany}

\date{\today}

\begin{abstract}
With the emergence of an information society, the idea of protecting sensitive data is steadily gaining importance. Conventional encryption methods may not be sufficient to guarantee data protection in the future. Quantum key distribution (QKD) is an emerging technology that exploits fundamental physical properties to guarantee perfect security in theory. However, it is not easy to ensure in practice that the implementations of QKD systems are exactly in line with the theoretical specifications. Such theory-practice deviations can open loopholes and compromise security. Several such loopholes have been discovered and investigated in the last decade. These activities have motivated the proposal and implementation of appropriate countermeasures, thereby preventing future attacks and enhancing the practical security of QKD. This article introduces the so-called field of quantum hacking by summarizing a variety of attacks and their prevention mechanisms.
\end{abstract}

%\pacs{03.67.Dd}% PACS, the Physics and Astronomy Classification Scheme.

\maketitle

%############################################################%
\section{Background}\label{Background}
%############################################################%
\subsection{The need for security}
The art of making and breaking secrets can be traced to the early civilizations~\cite{Singh1999}. The word \emph{crypto} for example has its origins in ancient Greek and means `hidden' or `secret'. The seeds of modern cryptology were sown around 1200 years ago by the Arabs, who invented systematic methods such as frequency analysis to unravel hidden messages. With the advent and rise of information and telecommunication technologies, most notably the Internet, the interest in cryptography has shot up exponentially in the last decades. Increasing amounts of text and voice records, e.g.~in emails and phone calls, are being cryptographically secured every day. This implies that the contents of the messages are (hopefully) made indecipherable to and unalterable by unauthorized parties --- often called adversaries --- and that the identities of the sender and the receiver are mutually attested. In particular, industry, banks, and governments strongly depend on such confidential communication.
\subsubsection{Cast of characters}
From a communication point of view, the most fundamental setting to study cryptography involves three entities: the message sender \emph{Alice}, the message receiver \emph{Bob}, and an adversary \emph{Eve}, who is not authorized to know the contents of these messages but is nonetheless (malevolent and) interested in doing so. Alice and Bob are connected together by one or more communication channels over which they exchange --- send and receive --- these messages. Eve can exert varying amounts of control on these channels. She may just passively eavesdrop, i.e. listen to the exchange and simply copy it as best as possible for analysis. Alternatively, she may actively tamper with the messages, i.e. modify the existing content and/or inject new messages. Finally, a technically-powerful Eve may simultaneously impersonate Alice (and communicate with Bob) and Bob (and communicate with Alice) successfully, thus learning all of their secrets.
%############################################################%
\subsubsection{Basic security goals}\label{Background:BasicGoals}
Cryptographic schemes (also called ciphers) in modern times follow the doctrine 
%(known by names such as Kerckhoff's principle~\cite{Kerckhoffs1883} or Shannon's maxim~\cite{Shannon1949}) 
that an adversary can/will always find out how the system or the algorithm works. The main implication of this doctrine is that all except one piece of information --- aptly called the \emph{private key} --- must be simply assumed to be known to the adversary. In a \emph{symmetric} cipher, the keys held by Alice and Bob are the same, while in an \emph{asymmetric} cipher, Alice and Bob each hold a key pair called public and private key. Either way, these ciphers fulfill three basic security requirements against the aforementioned actions of Eve:
\begin{enumerate}
\item \textbf{Confidentiality:} The message can be encrypted --- converted to a form from which Eve cannot derive any useful information about the original message --- before being sent on the channel. Only entities that possess the key(s) can correctly decrypt the encryption, thus ensuring that the message contents remain confidential to only Alice and Bob. 
\item \textbf{Integrity:} Upon receiving the message, Bob can get evidence if it was altered in transit. This step reveals any potential tampering by Eve, thereby assuring the integrity of the message. Depending on the desired application, integrity can either be assured in a standalone manner or along with confidentiality. 
\item \textbf{Authenticity:} During the entire communication, Alice/Bob have the confidence that the entity on the other end of the channel is Bob/Alice and not Eve. Just like for encryption, Alice and Bob can authenticate themselves using keys that are known to only them, denying Eve the opportunity to pose as Bob to Alice or vice-versa.
\end{enumerate}
For most communication-based applications these days, the task of authentication usually also addresses the requirement of integrity. If Alice's bit sequence has been altered by Eve, both integrity and authenticity will be violated. Authentication combined with encryption, aptly called \emph{authenticated encryption}, forms one of the most natural notions of information security~\cite{Bellare2000}.
%############################################################%
\subsection{Computational vs.\ Information-theoretic security}\label{Background:CompVsInfo}
\begin{figure}%[!t]
\centering
\includegraphics[width=0.38\textwidth]{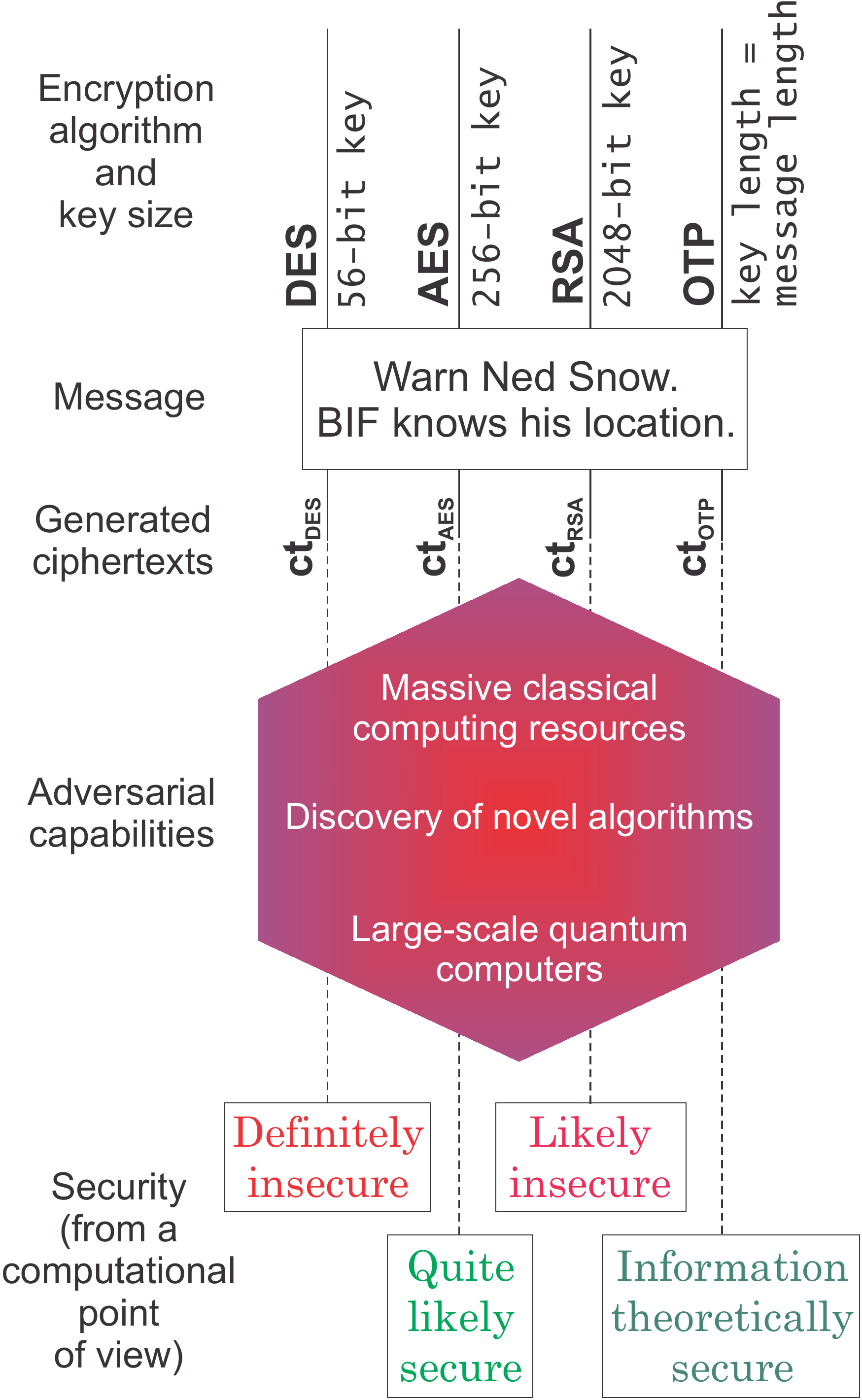}
\caption{(Color online) Computational versus Information-theoretic security. The four ciphers: data encryption standard (DES), advanced encryption standard (AES), RSA, and one time pad (OTP), can be considered to represent the contemporary cryptographic era and, possibly, its future. They feature varying key lengths and differ algorithmically. In this fictitous example, it is assumed that only the legitimate parties possess the keys for the four ciphers. An adversary captures the corresponding ciphertexts $\textbf{ct}_{\rm \textbf{X}}$s, with \textbf{X} denoting a cipher, and performs cryptanalysis with whatever abilities he/she has in order to know the original message. Depending on the cipher, the adversary may or may not be able to breach the cryptographic security today or in (near) future. Note that longer keys do not necessarily imply a higher level of security.\label{fcip21}}
\end{figure}

An obvious question to ask at this point is, `What guarantees the security of these ciphers?' In the modern context, perhaps the most basic requirement is that the keys must not be guessable for the adversary. This in turn implies that the keys must be sufficiently long and random as otherwise, the adversary may guess the correct key by means of an exhaustive search. Keys are typically realized as random bitstrings of a certain length. For instance, the advanced encryption standard (AES) cipher, which forms the workhorse of cryptographic security nowadays, typically employs a 256-bit key. To break a \emph{single} AES encryption, an exhaustive search would take $2^{256}>10^{75}$ steps, currently requiring billions of years with state-of-the-art ultra-massive computing resources. Conversely, a short key, such as in the data encryption standard (DES) cipher, which was employed until the early years of this century, can be guessed in a matter of days on a normal computer or hours on a cluster of computing devices. This is the reason why DES has officially been declared insecure, as also illustrated in Fig.~\ref{fcip21}.

AES and DES are examples of symmetric ciphers. Asymmetric ciphers, such as the RSA (named after its inventors Rivest, Shamir, and Adleman), typically use \emph{computational hardness} to make it difficult for the adversary to breach the security~\cite{Diffie1976,Rivest1978}. These ciphers are based on intractable mathematical problems that make the cryptanalysis --- recovery of the inputs from the output --- very difficult. To elaborate, RSA is based on the prime factorization problem which essentially states that it is computationally hard to find prime numbers $p$ and $q$, if \emph{only} their product $N (= p\cdot q)$ is known. Typical numbers involved in RSA problems these days are a few hundred digits long, translating to key lengths typically between $1024$ to $2048$ bits. 

But surprisingly, many of the mathematical intractabilities exist only because no \emph{efficient} methods of solving them (on classical computers) have been found up to date~\footnote{In fact, the proof for non-existence of such methods is related to one of the famous millennium problems.}. These include the discrete logarithm and the prime factorization problems that together form the foundations of several asymmetric ciphers in use today. It is clear that the discovery of new algorithms that can be efficiently operated on classical computers to solve these problems would put the security of such ciphers in jeopardy. Even otherwise, the security is at risk against an adversary in possession of unusually large computational resources that can run inefficient algorithms, yet solve the problems in a reasonable amount of time. Finally, efficient algorithms to crack these problems in the quantum domain are already known~\cite{Shor1994}. As soon as quantum computers of large enough scale become a reality, many well-known asymmetric ciphers would be easily broken. 
Figure~\ref{fcip21} depicts these scenarios against a message encrypted using RSA-2048 bit key.

The security (or lack of it) of DES, AES, and RSA can thus only be defined in a computational sense. In other words, given the current state and progress of classical and quantum computing, while DES/RSA are definitely/likely insecure, AES may be considered computationally secure from a practical perspective. There does exist another encryption mechanism known as the one time pad (OTP) that does not make any assumptions about the computational capabilities of the adversary. In other words, it guarantees information-theoretic security as also shown in Fig.~\ref{fcip21}.  
%############################################################%
\subsection{The problem of key exchange\label{Background:KeyEx}}
The theoretical origins of deploying the one time pad as a private key for encryption can be found in Shannon's seminal work~\cite{Shannon1949} while the first practical ideas for implementation are due to Vernam~\cite{vernam19}. Despite these developments being more than half a century old, one rarely hears of the OTP cipher in cryptographic installations around the world. As also shown in Fig.~\ref{fcip21}, this is due to the fact that the key/pad length must be at least as long as the message. In addition, the key must be refreshed for every new message. For typical communication bandwidths (of few Mbps) nowadays, these requirements imply that Alice and Bob should somehow be able to access a very large amount of private key material \emph{before} the actual confidential exchange. This can be quite expensive to satisfy for just two users; in a multiuser environment such as the Internet, the cost of assuring unique, random, and trustworthy keys to each user would scale up to prohibitive levels with just a handful of users. 

The alternative has been to use computationally-secure symmetric ciphers, such as AES, where just a few hundreds of bits of the private keys can be used for encrypting Gbits of message data. Nonetheless, the problem of how to securely distribute such private keys amongst Alice and Bob before they can communicate remains open. The current solution is to use asymmetric ciphers, such as RSA, for \emph{only} the key exchange. The security premise is based on the fact that efficient algorithms to break asymmetric ciphers have so far not been found. However, this is a \emph{ticking bomb} because as soon as an efficient algorithm is discovered, or in some cases, the availability of advanced quantum computers becomes a reality, the security provided by asymmetric ciphers will be in a major crisis. Compounding the worry is the fact that an adversary could simply be collecting and storing all ciphertexts today to be able to break their security in the future. 
Some data must remain secret for a long time, and in the modern information society, this includes not just military or financial secrets, but even data from individuals (for example, genetic codes may require protection for multiple generations).
%############################################################%
\subsection{(Post-)Quantum cryptography}
There are currently two approaches to tackle the above problem. The first, called quantum key distribution~\cite{Gisin2002,Scarani2009a}, addresses the problem of key distribution directly using principles of quantum physics. The other approach/field, called post-quantum cryptography, looks at the development of novel classical ciphers that would be invulnerable to quantum computers~\cite{Bernstein2008,Perlner2009}. Note that even though its definition allows it to be, quantum key distribution (QKD) was historically not considered a part of post-quantum cryptography (PQC). The European Telecommunication Standards Institute (ETSI) has recently taken cognizance of this by bringing both QKD and PQC under an umbrella term: quantum-safe cryptography~\cite{ETSI2014}.

The ideas that led to post-quantum cryptography were first conceived in the late 1970s~\cite{mceliece1978,lamport1979}. The first formal proposal of QKD was made in 1984 under the name quantum cryptography~\cite{Bennett1984} though the idea of quantum money~\cite{Wiesner1983}, dating back to late 1960s, is now recognized as a precursor of this proposal. Although, as of today, both QKD and PQC are conspicuous by their absence in realistic communication systems/networks (most notably the Internet), concerted efforts are being made to facilitate their integration in the existing infrastructure. In the following, we will concentrate on quantum key distribution.
%############################################################%
\section{Quantum key distribution (QKD)}
\subsection{Motivation: Optical communication and security}
In daily life, we usually deal with macroscopic objects that are too large to perceive any quantum effects. A light bulb switch, for example, can be either in the on-state or in the off-state, but not in both at the same time. These two states of the switch-light bulb system are of course clearly distinguishable as well. In quantum mechanical terms, they are orthogonal. One can then imagine Alice to use this lamp in order to transmit messages from her room to one of her neighbours Bob who lives somewhere in her line of sight. For that purpose, Alice would encode her message in bits: lamp on would mean Bit `1' and lamp off represent the Bit `0'. Bob simply has to write down the sequence of on/off states in order to obtain the message. This way, Alice and Bob effectively establish an optical communication link, resembling the beacon fires used in ancient times for transmitting signals along e.g. the Roman Limes or the Great Wall of China. 

However, if Alice and Bob want to exchange confidential messages via their line of sight link (formally called a \emph{channel}), they are in trouble. Since they have chosen orthogonal and therefore easily distinguishable light states for encoding their message, they cannot prevent Eve from also reading and decoding the message. 

Nowadays, Alice and Bob of course have more efficient ways of optical communication. Alice could use a laser followed by  a fast modulator and transmit her encoded light states to Bob via fibre-optic or atmospheric (line of sight) links. Bob would demodulate the states and detect them with fast photodiodes on which the impinging light states generate current pulses. However, the basic principle of using light states that are as orthogonal as possible remains the same in modern optical communication. This means that anyone else having access to the optical channel can read the messages. For instance, it recently became public that the British government communication headquarters (GCHQ) has been conducting fibre tapping activities on a large scale. For that purpose, GCHQ has installed intercepts at various landing points of undersea cables. Remarkably, fibre tapping itself does not actually require a sophisticated device since bending a fibre is sufficient to have some light leak out of it.

Clearly, not just in the `neighbour' scenario but also in the case of modern optical communication, it is feasible to \emph{eavesdrop} on an optical channel and obtain the message contents. In order to keep their communication secret, Alice and Bob must therefore encrypt their messages using a key as explained in subsection~\ref{Background:BasicGoals}. It turns out that they can use their optical communication system to share a random bit sequence known only to them -- essentially a secure symmetric key. But how do they get the key without sharing it with the eavesdropper? 

The solution, which consists of distribution of quantum states between Alice and Bob to generate a shared symmetric key, is called quantum key distribution (QKD). The optical channel used for the communication is called \emph{quantum channel} while the procedure is known as a \emph{protocol}. Notably, at the end of the protocol, Alice and Bob can find out whether the key distribution process was eavesdropped. The operation and security of QKD protocols are based on concepts in quantum and classical information processing~\cite{Gisin2002,Scarani2009a,Lo2014,niechu}. We discuss some of the key principles of QKD below, followed by a discussion on some common aspects of QKD protocols along with a description of BB84, the first and the most well-known QKD protocol~\cite{Bennett1984}.
%############################################################%
\subsection{Main principles \label{QKD:keyPrncpls}}
For easier comprehension we choose an example in which the polarisation direction of Alice's linearly polarized laser beam is to be determined. An oft-used method consists of rotating a polarizer around the beam axis until no intensity is observed on a photodetector placed after the polarizer. This means that the light is polarized exactly orthogonal to the polarizer axis. (Looking at a liquid crystal display with polarized sunglasses leads to the same effect). The equation
\begin{equation}
%\mathrm{cos}\left(\theta\right) = \frac{I}{I_0},
\cos\theta = \sqrt{\frac{I}{I_0}} \, ,
\label{Cos_Theta}
\end{equation}
also known as Malus' law, describes this phenomenon. Here $I_0$ and $I$ are the intensities before and after the polarizer, respectively, and $\theta$ is the searched value of the polarisation direction. 

Let us also imagine that Alice reduces the output power of her laser to a level where each signal state is a single photon pulse. Actually, attenuating a coherent laser does not exactly produce single-photon states but photon number superpositions that follow a Poisson distribution with the mean photon number $\mu < 1$. Such states are called weak coherent states and they can also be employed for QKD, as we shall elaborate later. For now, let us assume that Alice has very weak laser pulses with no more than one photon per pulse.
%############################################################%
\subsubsection{Non-orthogonality of quantum states \label{QKD:nonortho}}
A setup for testing Malus' law with such single photon pulses is possible since there are photodetectors that get triggered by only one photon (such events are often denoted by the term `clicks' of the detector). However, this photon is also destroyed by the measurement process and therefore can be measured only once. With only one single photon, we cannot measure the precise ratio $I/I_0$, because we have only two measurement outcomes: either the detector clicks or it does not. Notably, these discrete events happen with a $\cos^2\theta$ probability. 

\begin{figure}%[!t]
\centering
\includegraphics[width=0.38\textwidth]{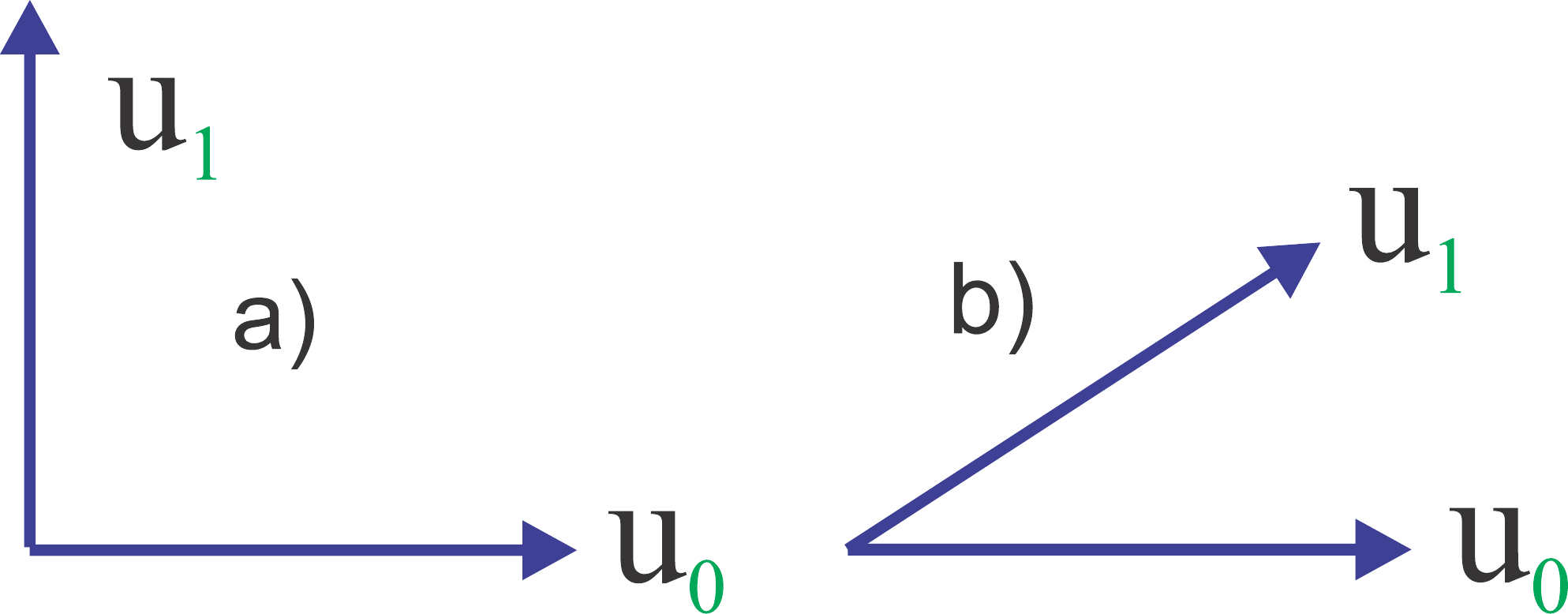}
\caption{Role of indistinguishability of non-orthogonal quantum states in QKD. a) The quantum states $u_0$ and $u_1$ encode bits `0' and `1' respectively. Since they are orthogonal, a measurement that can perfectly distinguish between them is possible. b) No such measurement exists for the case of non-orthogonal quantum states. The error in distinguishing increases as the angle between $u_0$ and $u_1$ decreases. \label{nonortho}}
\end{figure}

The only possible cases of deterministic measurements are when the polarisations of all single photons are aligned either exactly along or orthogonal to the polarizer axis. Figure~\ref{nonortho}(a) illustrates this situation. 
Two single photons with unknown linear polarisations that are not orthogonal to each other, as depicted in Fig.~\ref{nonortho}(b) for example, cannot be precisely discerned~\cite{niechu}. In other words, bits `0' and `1' encoded by these non-orthogonal polarisations cannot be decoded by any entity including Eve without making errors. In quantum mechanical parlance, Eve's interaction \emph{disturbs the quantum state} and this disturbance can both be detected and quantified. For that, Alice and Bob can later exchange some other classical information (about their preparation and measurement methods for each of the single photons). Crucially, if the amount of disturbance is below a certain level, Alice and Bob can classically distill a shorter bit string --- the secret key --- which is identical for Alice and Bob and on which Eve has no information.
%############################################################%
\subsubsection{No-cloning theorem\label{No_cloning_theorem}}
A classical copy machine produces duplicates by scanning the original which corresponds to a measurement process. As discussed above, such a process is not possible with \emph{unknown} quantum states since in general, a single measurement does not reveal the full information of the actual state. It can also be proven in general that quantum mechanics allows no other method to produce exact copies of unknown quantum states. This is known as the no-cloning theorem~\cite{Wootters1982}. 

For QKD, the no-cloning theorem is essential since it forbids an eavesdropper to exactly copy the quantum signal. Nonetheless, approximate or imperfect copies of the unknown quantum state can be produced by cloning machines~\cite{Buzek96}. An eavesdropping strategy based on cloning may therefore allow Eve to gain some information about the key without introducing any significant disturbances. To obtain a perfectly-secure key, Alice and Bob need to erase this partial information of Eve beforehand. 
%############################################################%
\subsubsection{Authenticated channel}
Apart from the quantum channel, Alice and Bob also need a classical communication channel for distilling a secret key out of their quantum state exchange. In contrast to the quantum channel, the information exchange on this classical channel can be public in the sense that Eve can listen to all messages of Alice and Bob. However, Eve must not be able to tamper with these classical messages since otherwise, she can perform the so-called man-in-the-middle attack in which she impersonates Bob for Alice and vice-versa allowing her to gain knowledge of the full key. To prevent such an attack, Alice and Bob must be authenticated to each other. This implies that a fully functioning QKD system provides authenticated encryption; see subsection~\ref{Background:BasicGoals}.

There are symmetric algorithms that provide information-theoretic security for authentication~\cite{Carter1979,Wegman1981}. During the operation of the QKD protocol, Alice and Bob can use a  small amount of the distilled key from the previous run for authentication purposes. However, this poses a difficulty for the very first key exchange. To elaborate, Alice and Bob would need a pre-shared key before they have initiated the key exchange process. Due to this, QKD is more accurately described as a key growing (instead of generation) process: an initially small key is grown to arbitrary length. 

A practical solution to this initial problem, also mentioned in subsection~\ref{Background:KeyEx}, would be to use asymmetric key ciphers only for the initial step. Notably, if Eve cannot break the asymmetric key during the short moment of the very first quantum key exchange, the system can operate securely. This is because all further authentication steps would be performed by symmetric algorithms that use a part of the secure key obtained from the QKD protocol~\cite{Stebila10}. 
%############################################################%
\subsection{QKD protocols \label{QKD:qkdprots}}
A QKD protocol describes the procedures and operations that Alice and Bob perform in order to generate a secret shared key sequence. A typical QKD protocol consists of two main steps: 
\begin{enumerate}
\item Quantum states are generated, transmitted through a quantum channel and measured. At the end of this exchange, Alice and Bob have bit sequences from which a shorter but correlated bit string could be extracted. Note that Eve can have partial knowledge of the extracted string.
\item In order to distill a secret, shared key out of their bit sequences, Alice and Bob perform classical data processing, typically consisting of sifting, error correction and privacy amplification. We will explain these terms in the coming pages.
\end{enumerate}
For the quantum states, it is crucial that their quantum properties are conserved (at least partially) after having propagated through the quantum channel. Otherwise Eve's attacks could go undiscovered and the security premises of QKD would not be valid anymore. Conservation of quantum properties means that the attenuation and noise in the quantum channel may not exceed certain levels. In practice, a variety of quantum-optical states, such as single-photon states, coherent and squeezed states, as well as entangled states have been tested for suitability in quantum channels implemented by optical fibres and free space links. The quantum information itself can be encoded in physical properties such as polarisation, amplitude and phase quadratures or time bins. The quantum state measurement strategy is naturally decided by the details of the quantum state preparation. For instance, polarisation-encoded single-photon states may be decoded using a polarizer followed by single photon detector(s), coherent states encoded in quadratures may be decoded using phase-sensitive homodyne detection~\cite{Gisin2002,Scarani2009a,Ralph1999}.
%############################################################%
\subsubsection{Example of a QKD protocol: BB84 \label{QKD:rawkey}}
While the details of the quantum state preparation and measurement obviously depend on the specifics of a QKD protocol, a fairly representative explanation can be given through the BB84 protocol~\cite{Bennett1984}. This protocol was proposed at a conference in Bangalore, India in 1984 by Charles Bennett and Gilles Brassard. We consider here a BB84 implementation with quantum information encoded in one of four different polarisation states of single photons. 

\begin{figure}%[!htb]
\centering
\includegraphics[width=0.3\textwidth]{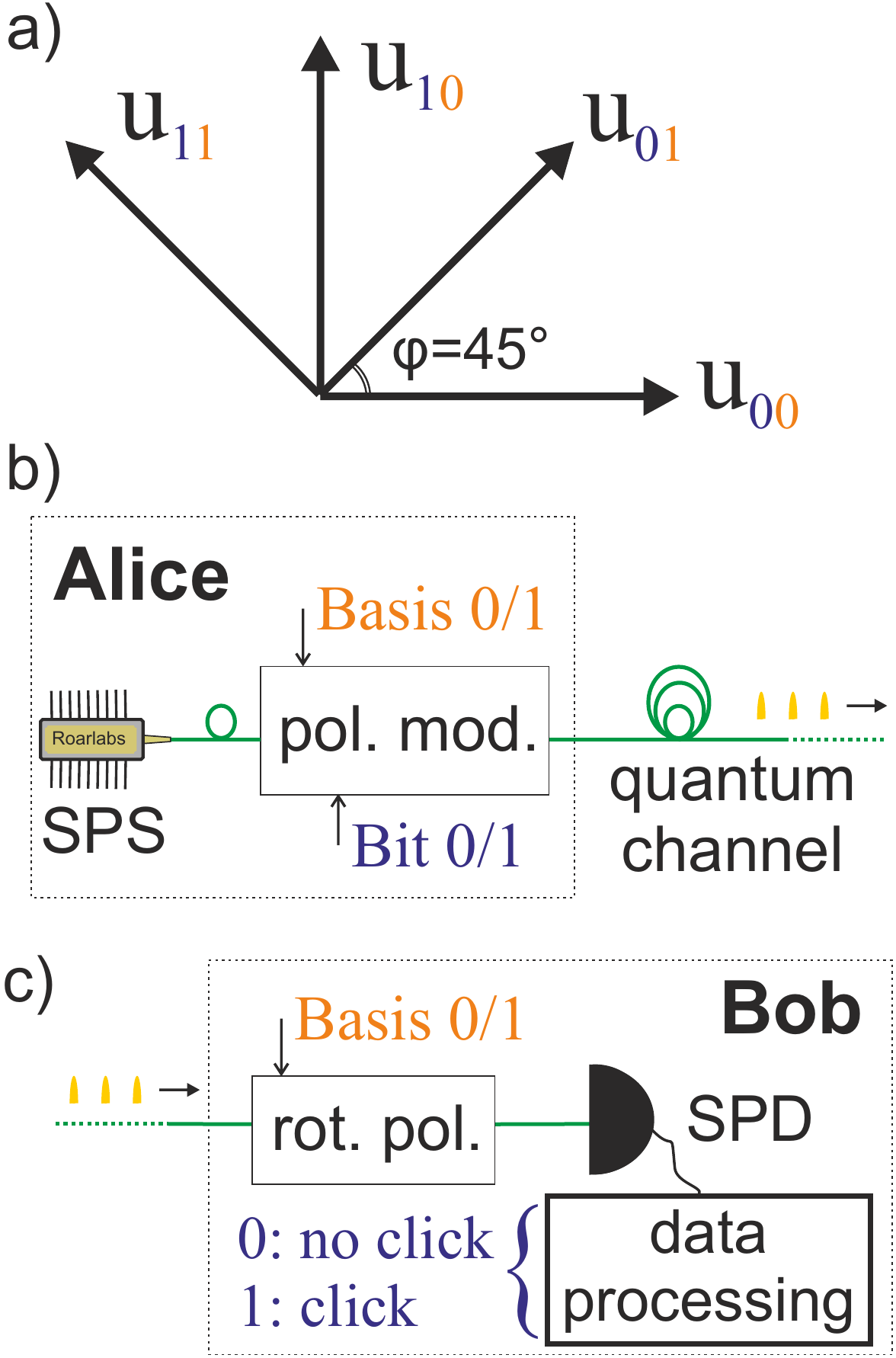}
\caption{(Color online) Polarisation coding in BB84. a) The BB84 state alphabet can be represented by two pairs of mutually orthogonal vectors; see Fig.~\ref{nonortho}. In order to relate to polarisation, the mapping $u_{00} \equiv \ket{H}$, $u_{01} \equiv \ket{D}$, $u_{10} \equiv \ket{V}$, and $u_{11} \equiv \ket{A}$ may be used. b) Alice can prepare the BB84 quantum states by modulating the polarisation of single photons. c) Bob can use a fast rotating polariser (rot. pol.) to measure the polarisation of the incoming quantum states and obtain the bit of Alice dependent on the detection outcome. SPS/SPD: single photon source/detector.
\label{fBB84AliceBob}}
\end{figure}

For this purpose, Alice randomly switches between a pair of mutually unbiased bases: in the Basis 0, denoted by the orange/second subscript in Fig.~\ref{fBB84AliceBob}a, the bit $b_A=1$ is encoded as a vertically polarized photon, $b_A=0$ as a horizontally polarized photon (bits are denoted by the blue/first subscript). 
Similarly, in the Basis 1, the bits $0$ and $1$ are encoded as diagonally and anti-diagonally polarized photons. Alice can generate these different polarisation states using four differently oriented single photon sources (SPSs) or by modulating the polarisation of a single SPS, as shown in Fig.~\ref{fBB84AliceBob}b.

After propagation through the channel, Bob measures each quantum state, also by randomly switching between one of the two bases followed by detecting the photon. He can switch the basis by rotating a polariser to either vertical or $45^{\circ}$ direction, corresponding to Basis 0 or 1 (see Fig.~\ref{fBB84AliceBob}c). In the cases where Bob's basis choice matches with Alice's, a `Yes' or `No' detector can measure a conclusive result: when the detector clicks, bit $b_B = b_A = 1$ (Yes), when there is no click, bit $b_B = b_A = 0$ (No). In the cases where Bob chooses a different basis to Alice, his measurement results will be random and inconclusive (sometimes Yes, sometimes No). 

To begin the classical processing, Alice and Bob run through their entire bitstreams, sifting all inconclusive events. For this, either of them discloses the bases (but not the bit values) on the authenticated channel. Both parties then learn of the slots where their bases had matched; the corresponding bits together make the `raw key'. Assuming the (unrealistic) case of a loss-less and noise-free quantum channel, Alice and Bob should possess identical raw keys after sifting.
%############################################################%
\subsubsection{How to catch the eavesdropper \label{QKD:CatchEve}}
The non-orthogonality principle and the no-cloning theorem, explained in subsection~\ref{QKD:keyPrncpls}, ensure that Eve cannot obtain information of the raw key without introducing errors in the bits that Alice and Bob obtain at the end of sifting. In fact, as the next stage of the protocol, Alice and Bob publicly disclose a small and random portion of their respective raw keys and check for errors ($b_A \neq b_B$ in a given slot). Performing this allows them to numerically estimate the rate at which errors actually occur in the raw key. This yields the quantum bit error rate (QBER)~\cite{Townsend1998}, which is one of the most important parameters for gauging the extent of eavesdropping. 

The QBER also determines the course of the rest of the protocol; for instance, the next step of \emph{error correction}, which is necessary to remove the mismatches in the raw keys, is strongly dependent on the value of the QBER. More importantly and as we shall explain below (see also subsection~\ref{qh:Conds}), if the QBER is above some critical value, no secure key can be distilled. In such a case, Alice and Bob must either resort to another quantum channel or try again later. Although no doubt an undesirable situation, the fact that Alice and Bob realize that the communication channels are not safe provides them the chance to protect their secrets. At present, no classical cryptographic mechanism that provides such a functionality is known. 

The error correction is followed by a distillation procedure in which a shorter but highly secure key is distilled from the longer but perhaps insecure raw key. This stage is known as \emph{privacy amplification}~\cite{Bennett1988,Renner2005} since it increases the confidentiality of the key that Alice and Bob shared. At the end of this stage, the probability that Eve has any information of the secret key is reduced to below some requisite threshold value.%, usually denoted by $\epsilon$; typical QKD systems aim for $\epsilon \sim 10^{-9}$ nowadays. 
%############################################################%
\subsection{Security proofs\label{SecProofs}}
In theory, quantum key distribution can guarantee information-theoretic security of the message encrypted by the exchanged key --- employed as one time pad --- under certain conditions. These conditions are elaborated in quantum-mechanical security proofs that theoretically model the actions of all the involved parties (Alice, Bob, and Eve) and evaluate limits under which privacy amplification is successful in nullifying Eve's information of the secret key. 

Several different approaches for constructing security proofs (for even the same QKD protocol) are known as of today~\cite{Lo1999,Shor2000b,Lutkenhaus2000,Renner2005b,Kraus2005,Branciard2005,Inamori2007} but at the heart of the majority is a limiting value $\rm Q_{L}$ of the QBER. As also explained before, this is because the eavesdropper's actions perturb the quantum states, causing errors in the measurement stage. The amount of errors can be quantified by Alice and Bob to obtain bounds on Eve's information gain. Consequently, these proofs also provide a security assurance --- in the form of bounds on the obtainable secret key rates --- at the end of the protocol. Roughly speaking, if the incurred QBER value $q > \rm Q_{L}$ the security of the distilled key may no longer be guaranteed.

The process of theorizing a security proof for numerically evaluating $\rm Q_{L}$ involves making assumptions about the physical devices of Alice and Bob, the quantum channel connecting them, the optimal actions of Eve, etc. As an example, many well-established security proofs calculate a value of $\rm Q_{L} \approx 11.0\%$ for the BB84 protocol. In these proofs, the state preparation in Alice and state measurement in Bob are assumed to be ideal. To elaborate, pure single photons are perfectly encoded by Alice while Bob decodes each of the arriving single photons with noiseless detectors. The subsystems of Alice and Bob are also assumed to be perfectly aligned. Eve is assumed to have no physical access to Alice and Bob but the channels that connect them are her territories. On the quantum channel, she can interact with the quantum signals traversing from Alice to Bob. On the authenticated channel, Eve cannot modify the exchanged messages but may read them without paying any penalty. %($\eta = 1$) ($d = 0$)
%############################################################%
\subsubsection*{Classification of attacks \label{SecProofs:ThAttcks}}
There are an inexhaustible number of strategies that Eve may employ to attack the QKD system. The final objective is to obtain information about the secret key without alerting Alice and Bob. Let us explain one of the most well-known strategies called the intercept and resend attack (IRA). Eve intercepts and measures each of the signals herself and according to the measurement result, re-prepares new signals to send to Bob. Assuming Alice and Bob operate the BB84 protocol, Eve's measurements (and hence the re-prepared states) can be correct in only half the total number of instances. Out of the remaining half, Bob's measurements would be erroneous with a probability of $1/2$ since Eve's choice of bases did not coincide with that of Bob and Alice. In other words, Alice and Bob would notice errors in around a fourth of all the cases pointing to $q \sim 25.0\%$ which is clearly much higher than the tolerable limit of $\rm Q_{L} \approx 11.0\%$. However, if Eve performs IRA only on a fraction $f < 11/25$ of all the quantum signals, the QBER incurred by Alice and Bob may not cross the abort threshold; albeit Eve's knowledge of the raw key is also reduced from a factor of $0.5$ to $(f*0.5 = )0.22$. Alice and Bob must then make sure that this partial information is removed in the privacy amplification step. 

The intercept and resend attack strategy belongs to the class of so-called \emph{individual attacks}~\cite{Gisin2002}. This class caters to attacks in which Eve individually interacts with the quantum signals that are enroute from Alice to Bob. Individual attacks that perform better than an IRA are already known explicitly for many QKD protocols~\cite{Fuchs1997,Branciard2005,Renner2005b,Niederberger2005}. This implies that at some $q < \rm Q_{L}$, Eve's knowledge of the raw key can be higher than that obtained via IRA.

The amount of privacy amplification is usually evaluated in the so-called \emph{collective attacks} scenario by most security proofs~\cite{Biham1997,Renner2005,Scarani2009a}. This class of attacks imposes less restrictions on Eve's actions than individual attacks, thus making the security proof more general and stronger. To make things more precise, the security of a QKD protocol calculated under the assumption of collective attacks automatically insures the QKD system (operating that QKD protocol) against the best individual attacks. 
Nonetheless, collective attacks are still limited to specific attack strategies, in contrast to \emph{coherent attacks} that account for any possible attack strategies limited only by quantum mechanics. However, and quite remarkably so, security against collective attacks also implies security against coherent attacks under some reasonable assumptions~\cite{Renner2005}. 
%############################################################%
\section{Quantum hacking \label{qh}}
So far we discussed various aspects of quantum key distribution assuming an ideal world. In real life, the operation of a QKD protocol deviates from the ideal due to imperfect components that make up the physical QKD system. Such deviations can have major implications not just for the design and performance of the system but also the security guarantees against attacks~\cite{Scarani2014}. 

The field of quantum hacking investigates theory-practice deviations that specifically result in security loopholes. Given a potential deviation, e.g. due to the behaviour of a specific set of components, the first step is to experimentally confirm the existence of the loophole in the QKD implementation. This is followed by quantifying the impact of the loophole through attack simulations or further experiments. One tests the design of a new QKD system by engineering practical attacks based on known weaknesses and existing loopholes. 

Despite being a fairly young field, quantum hacking has caught the interest of both the research community and (scientific) media: more than half-a-dozen research labs across the world are known to be actively engaged in this area, the number of publications in peer-reviewed journals and respectable magazines on this topic has skyrocketed, and `live hacking' demonstrations have been performed at international conferences. 
Nonetheless, it must be emphasized that the intention of quantum hacking activities is to improve the security obtainable by practical QKD implementations. In fact, a complete implementation of the eavesdropping system --- demonstration of the attack on fully-functional Alice and Bob connected by `proper' quantum and authenticated channels --- is rarely performed: most quantum hacking attempts display the insecurity of the QKD system only in a \emph{proof-of-principle} manner. 
The belief is that bugs and loopholes are a part and parcel of any developing technology and practical QKD is no different in that regard. A tight scrutiny can therefore ensure the removal of vulnerabilities and patching of loopholes to guarantee reasonable and acceptable level of security of messages encrypted by QKD systems in the future. 
%############################################################%
\subsection{Realistic QKD}
We first discuss different aspects of physical QKD systems to gain an insight into the causes and effects of problems associated with the field of realistic QKD. We do so by considering the state preparation and measurement example from subsection~\ref{QKD:qkdprots} in presence of channels, sources, and detectors that are employed in practical QKD systems.  
%############################################################%
\subsubsection{Realistic channels}
Absorption and scattering in the channel lead to photon loss due to which not all quantum states from Alice reach Bob in practice. Furthermore, the ones that do arrive may be decoded incorrectly because of the noise induced by the channel. (Note that Eve's actions on the quantum channel can also certainly lead to losses apart from an increase in the noise, however, here we focus on an implementation without Eve). This calls for a characterization of the loss and noise properties of the channel. 
A frequently-used parameter in that regard is the transmittance $T$ which may be understood as the fraction of the number of photons from Alice that are received by Bob. Any realistic channel would have $T<1$. 
%############################################################%
\subsubsection{Realistic sources and detectors\label{realSrcDets}}
Apart from the problems with realistic channels, realistic sources and detectors also do not generate and detect photons in a perfect manner. Most single photon sources (SPSs) show a probability distribution in their photon number $n$, featuring non-zero values at $n=0$ and $n \geq 2$. In fact, the most popular source used for mimicking SPS in typical QKD implementations are attenuated lasers. As mentioned before, the corresponding quantum states (called weak coherent states) are characterized by a Poissonian distribution with the mean photon number $\mu$. Due to security reasons that will be explained later, attenuated lasers in Alice employ $\mu < 1$ implying the output quantum signal is mostly $\ket{n=0}$. The single detector configuration, shown in Fig.~\ref{fBB84AliceBob}c, is therefore completely impractical because the absence of a click (that would suggest bit 0) may actually have occurred because no photon was generated by Alice or the generated photon was lost in the channel due to attenuation. 

\begin{figure}
\centering
\includegraphics[width=0.4\textwidth]{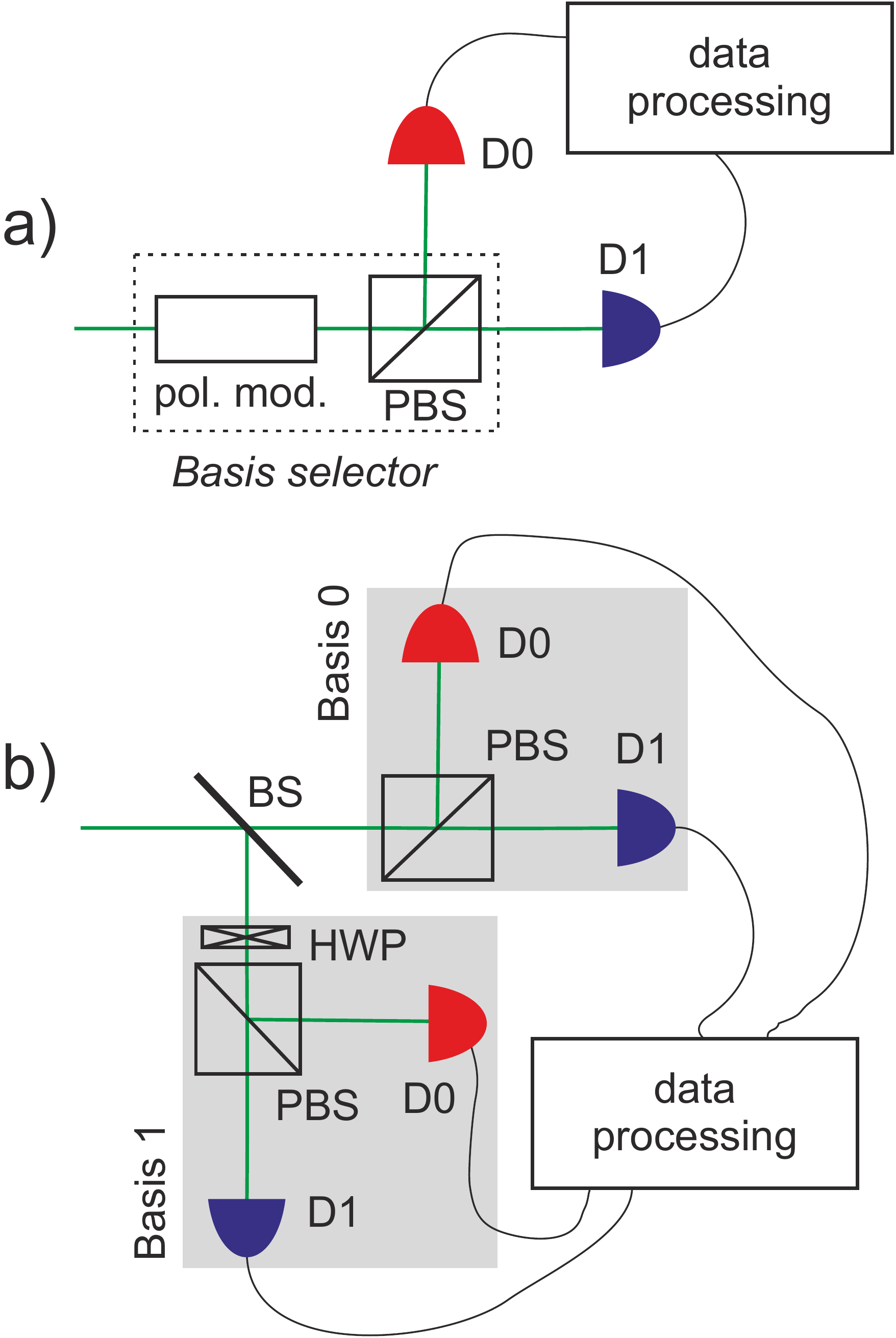}
\caption{Two practical measurement setups for polarisation-coded BB84. a) Two detectors are used in order to reduce inconclusive measurement outcomes due to losses and noise. b) Using four detectors makes the polarisation modulator dispensable. The basis choice is then mediated passively by a non-polarising beam splitter (BS). HWP: half wave plate, pol. mod.: polarisation modulator, PBS: polarising beam splitter, D0 and D1: detectors}
\label{BB84BobSetups}
\end{figure}

A common practice is therefore to employ two detectors (labeled D0 and D1; see Fig.~\ref{BB84BobSetups}a) to measure orthogonal polarisations. 
Inconclusive events when neither of the two detectors clicks can now be discarded safely. Note that the basis choice can be done actively, such as by using a polarisation modulator. Alternatively, the choice between Basis 0 or Basis 1 is passively mediated by a 50/50 beam splitter with the quadruple detector assembly, as depicted in Fig.~\ref{BB84BobSetups}b. (The half wave plate rotates one orthogonal polarisation pair into the other). Due to the beam splitter, this methods leads to 50\% of additional loss as compared to the scheme with polarisation modulator.

Just like channels, the loss and noise properties of the detectors must also be characterized. The probability of detecting a single photon is given by a detection efficiency $\eta < 1$. Many realistic detectors also suffer from the so-called dark noise (usually denoted by a dark click probability $d > 0$) which means that they sometimes produce clicks even in the absence of photons. 
%############################################################%
\subsubsection*{Single photon avalanche diodes (SPADs)}
To explain these concepts further, we take up the example of single photon avalanche diodes (SPADs) that are the workhorse detectors for quantum states containing single/few photons~\cite{Hadfield2009}. This minor detour is also necessitated by the fact that a major part of the discussion presented in section~\ref{Lnaip} requires a basic understanding of SPADs. 

\begin{figure}[!b]
\centering
\includegraphics[width=0.4\textwidth]{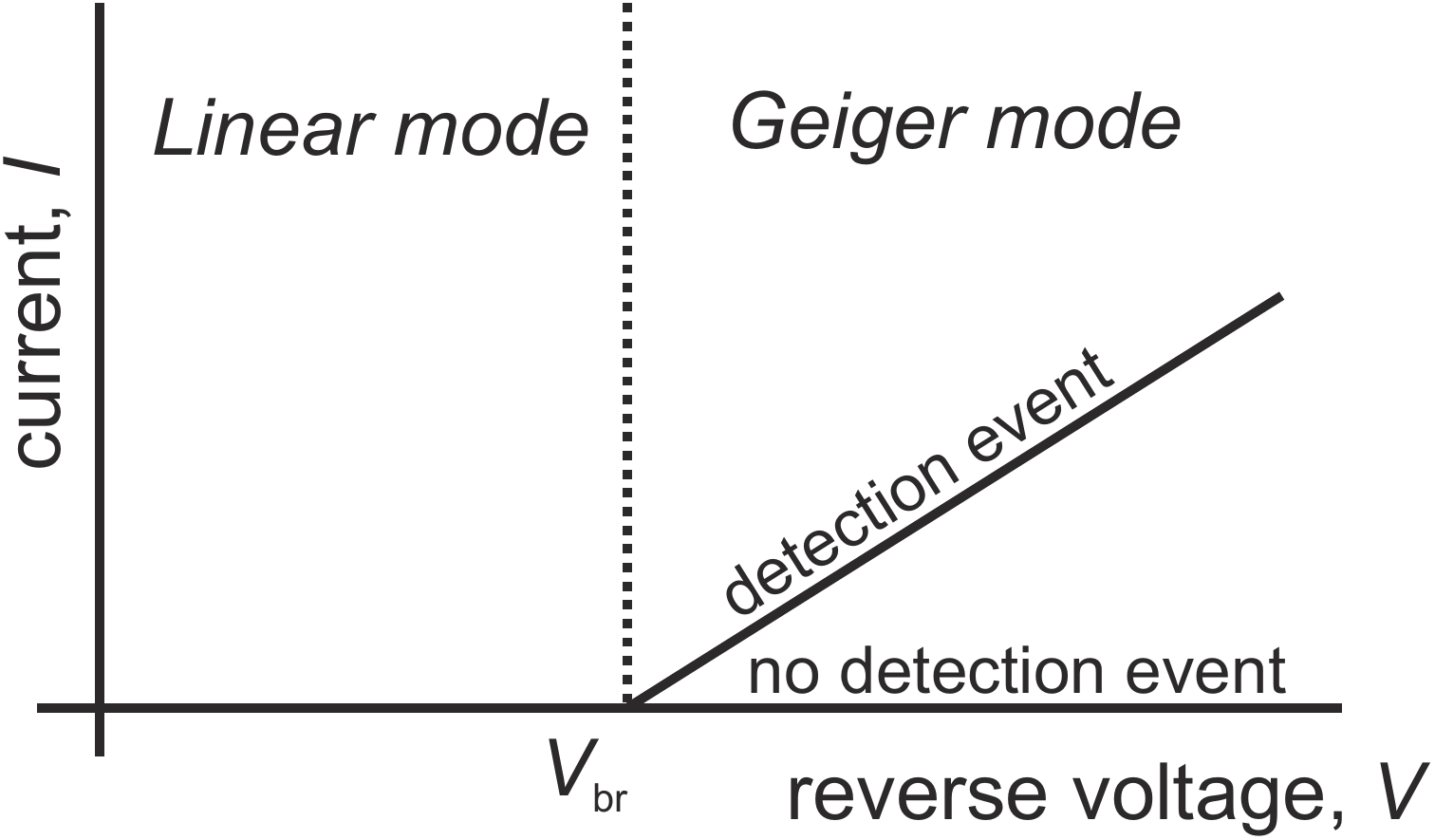}
\caption{Current-voltage characteristics of a single photon avalanche diode (SPAD). To detect a single photon, the SPAD is operated in the Geiger mode where a voltage larger than the breakdown voltage $V_{br}$ is applied across the SPAD. In this regime, a single photon can generate an avalanche of electron-hole pairs. This avalanche produces a measurable current through the diode which indicates the detection of the photon.}
\label{APD_I_V_Diagram}
\end{figure}

Figure~\ref{APD_I_V_Diagram} shows the working principle of an SPAD via its current-voltage relationship. One can observe a bifurcation in the current-voltage diagram above the breakdown voltage $V_{br}$ of the diode. The rising current in the upper branch of the bifurcation stems from an avalanche that, in the nominal case, has been triggered by a single photon. However, in most SPADs, this event only happens with a non-unity probability $\eta < 1$. In other words, it might happen that the impinging photon fails to excite an electron-hole pair. On the other hand, these carriers may be generated due to thermal excitations inside the SPAD. In the event they result in an avalanche, the SPAD can register a dark click (denoted by a probability $d>0$) which cannot be distinguished from photon clicks.
Values such as $\eta \approx 0.2-0.3$ and $d < 10^{-5}$ are typically obtained for SPADs used in QKD systems these days~\cite{Scarani2009a,Hadfield2009,Itzler2011}.
%############################################################%
\subsubsection{Finite QBER without Eve}
In addition to dark noise, wrong detection events can also be caused by imperfect alignment of some optical components. A misaligned modulator in Fig.~\ref{BB84BobSetups}a, for example, would lead to erroneous detections (also see Equation~\ref{Cos_Theta}). Furthermore, stray light can lead to false clicks since SPDs cannot distinguish between signal photons and stray light.

To sum up, errors --- apart from stemming from Eve's actions --- in the shared key can also happen because of channel noise, dark click probability, optical misalignment within Bob and Alice, etc. Therefore, a finite QBER is obtained even without the actions of the adversary. Furthermore, channel losses and noise generally grow as the distance between Alice and Bob increases. Due to this, after a certain channel length, the QBER arising from these various imperfections and inefficiencies may easily exceed the critical value $\rm Q_{L}$ (see subsection~\ref{SecProofs}). 
%############################################################%
\subsubsection{Industrial implementations}
Many renowned institutes and corporate establishments actively pursue research in quantum technologies today. Startup firms such as ID~Quantique~\cite{idq}, one of the pioneers of practical QKD systems, have been selling commercial equipment for a decade now. Their approach blends conventional and quantum encryption: keys obtained independently via RSA and QKD are combined together for 256-bit AES encryption; see subsection~\ref{Background:CompVsInfo}. They also collaborate with research institutes: in fact, many attacks discussed in section~\ref{Lnaip} were performed on their research platform `Clavis2'~\cite{clavis2DS}. In the forthcoming section, we shall elaborate on these attacks. 
%############################################################%
\subsection{Imperfections and assumptions\label{ThryPracDevs}}
In the previous pages, the imperfect behaviour of light sources and detectors employed in QKD implementations was discussed. For instance, while realistic single photon sources may sometimes produce zero or multiple photons ($\ket{n=0}$ or $\ket{n>1}$ states, respectively), detectors may sometimes not only fail to detect a photon but also exhibit false detection events (efficiency $\eta < 1$ and dark noise level $d > 0$, respectively). Such intrinsic limitations have a major impact on the operation of the QKD system. For instance, dark noise results in a finite QBER even without Eve's presence in the quantum channel. 

Furthermore, the theoretical model considered by the security proofs (see subsection~\ref{SecProofs}) may make seemingly-harmless assumptions that can however not be validated, even in principle. For instance, many of the early security proofs implicitly assumed an asymptotic operation of the QKD protocol, i.e. Alice and Bob were assumed to exchange infinitely long keys. Only recently, the issue of non-infinite length keys was investigated and the insecurity arising from such `finite size' effects quantified~\cite{Inamori2007,Scarani2008a}. 

Imperfections in the QKD system hardware and insufficient/unverified assumptions in the security proof therefore lead to deviations between the theoretical model and the practical implementation. These theory-practice deviations often open loopholes in the security framework, rendering the QKD system vulnerable to quantum hacking. In other words, Eve could manipulate the operation of the QKD protocol using suitable attack strategies that violate the guarantees of the security proofs without Alice's and Bob's realization.
%############################################################%
\subsubsection{Side channels\label{schs}}
The nature of Eve's action depends on the loophole. For instance, Eve can actively manipulate the signals travelling from Alice to Bob on the quantum channel. Alternatively, she injects some radiation in Alice's and Bob's subsystems with the purpose of modifying the operation of a specific set of components or obtaining information about their settings. Such attacks generally carry the risk of being discovered. 

In contrast, Eve may also perform certain measurements that do not directly affect Alice or Bob. To be more precise, Eve does not interact with either the physical devices or the quantum signals. These attacks usually happen if and when there is a (possibly inadvertent) leakage of information from the QKD devices. The term `side channel' attacks is sometimes used in such contexts: one may interpret the encoding/decoding of information taking place on channels that exist beside the quantum and authenticated channels. For instance, imagine the sending device of Alice: a metal box that fits inside a standard 19'' rack. The box features an LED that is supposed to indicate the power status of the device. However, due to bad engineering, the LED's driving current becomes electronically coupled to the electro-optical modulator used for preparing the quantum state in a basis otherwise unknown to Eve. The impact is that the LED flashes depending on the state of the modulator, thus giving away the raw key of Alice to an eavesdropper just observing the device from a distance. 

In general, whether a side channel assists Eve depends on the precise type and quality of the information available through it. The main point though is that Eve can avoid alerting Alice and Bob of her eavesdropping actions in this case.
%############################################################%
\subsection{Conditions for successful breach\label{qh:Conds}}
A quantum hacking attempt is successful if it can be proven that Eve gains information of a non-negligible amount of the final secret key extracted by Alice and Bob without alerting them. The latter may be elaborated by saying that during Eve's attack, 
\begin{enumerate}
	\item the incurred QBER must not cross the abort threshold ($q\leq \rm Q_{abort}$), and
	\item the fluctuation in the value of channel transmittance observed by Alice and Bob must remain within tolerable limits ($\delta < \delta_{\rm abort}$). 
\end{enumerate} 
The precise values of $\rm Q_{abort}$ and $\delta_{\rm abort}$ hardwired in the physical QKD implementation ought to depend on the QKD protocol and the security proof it follows. The QKD device manufacturer may however choose more conservative values. For instance, the research platform `Clavis2' from ID Quantique features $\rm Q_{abort} \approx 8\% < Q_{L}$ and $\delta_{\rm abort} = 0.15$ for the BB84 protocol~\cite{Jain2011,Jain2014}. 

In the next section, we discuss several quantum hacking attempts that have been devised on a variety of QKD implementations in the last decade. We also analyze their performance with respect to the aforementioned conditions. Note that with the possible exception of the work presented in Ref.~\cite{Gerhardt2011}, all attacks were performed in a proof-of-principle manner.
%############################################################%
\section{Loopholes/Attacks in practice\label{Lnaip}}
In this section, we present the main ideas of several known attacks on practical QKD. A great deal of them exploit (natural) limitations in the single photon detectors and/or imperfections in their optoelectronic interfaces. The most commonly-discussed detector type in the next pages shall be the gated~\footnote{A train of narrow voltage pulses applied on the diode so that it switches between the linear mode and Geiger mode are called \emph{gates}. Refer subsection~\ref{realSrcDets} and Fig.~\ref{APD_I_V_Diagram} for details on these modes.} single photon avalanche diode (SPAD). While SPADs have been the workhorse detectors of single photons~\cite{Hadfield2009}, unfortunately, they have also proven to be one of the biggest source of vulnerabilities when it comes to practical security.
%############################################################%
\subsection{Faked-state attacks\label{Lnaip:fsas}}
A majority of attacks on single photon detectors in the last few years were based on the concept of \emph{faked states} of light~\cite{Makarov2005a}. These states are specially-crafted optical signals prepared by Eve and sent into Bob to control his detection outcomes in a manner dictated by her. 
\begin{figure}%[!htb]
\centering
\includegraphics[width=0.5\textwidth]{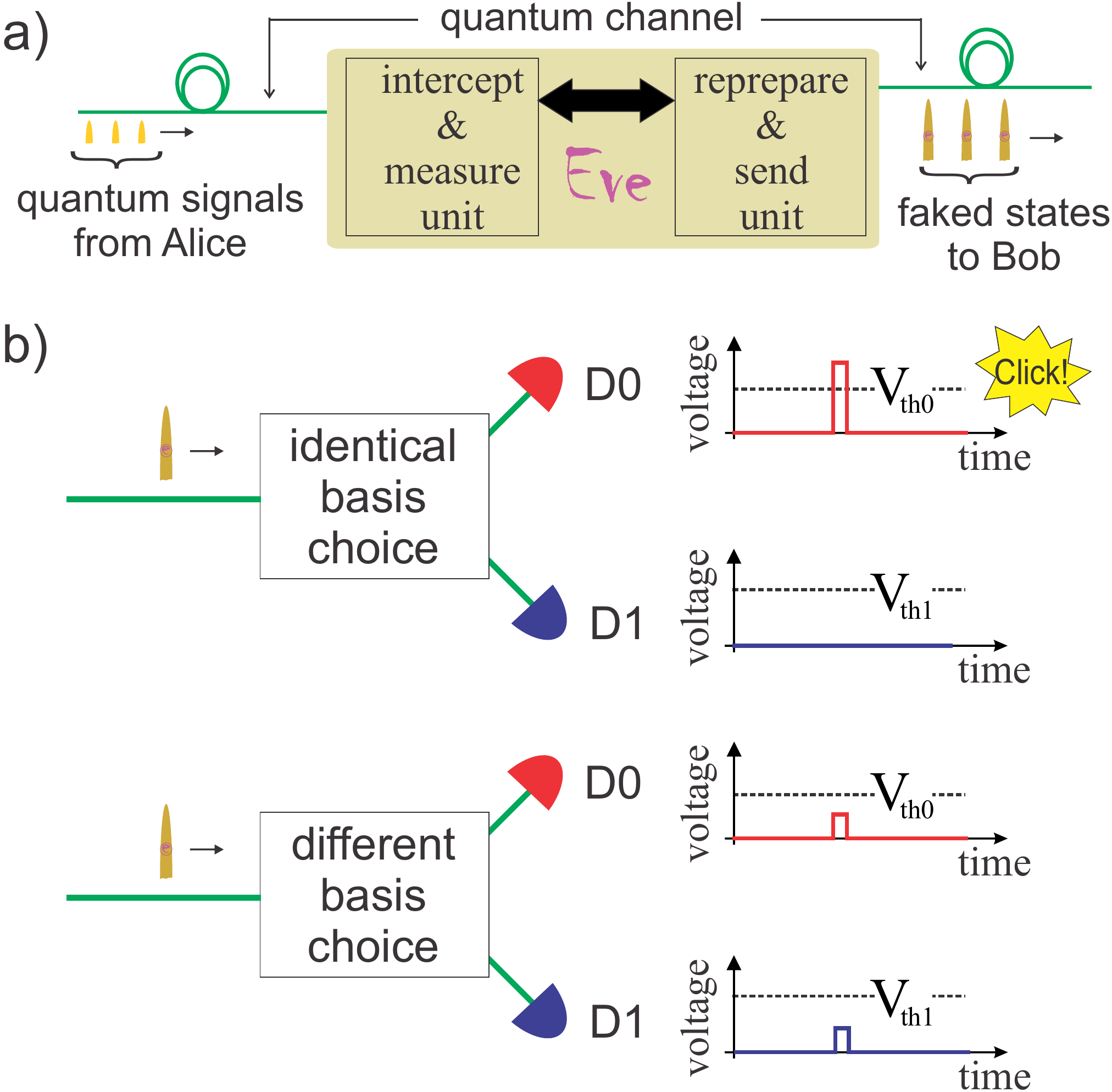}
\caption{(Color online) Basic principles of faked-state attacks against BB84. As discussed in subsection~\ref{SecProofs}, Eve intercepts and measures the quantum signals from Alice in a randomly-chosen basis. She prepares new states of light to fake detection events in Bob in a controllable manner. b) If the basis chosen by Bob is identical to that of Eve, the optical power of Eve's faked state is detected fully by one of the two detectors. The resulting voltage pulse crosses the `click' threshold, given by $V_{\rm thj}$ (for $j=0$ or $1$), and Bob obtains the bit corresponding to the detector that clicked as the measurement outcome. On the other hand, if the bases are different, the optical power is equally split across the detectors. The resulting voltage pulses in both D0 and D1 are below the respective thresholds $V_{\rm th0}$ and $V_{\rm th1}$, and no measurement outcome is recorded. 
\label{ffsa}}
\end{figure}
As shown in Fig.~\ref{ffsa}a, a faked-state attack (FSA) is implemented in the style of an intercept and resend attack (IRA), and targets detector imperfections to perform better than a conventional IRA. 
To be more specific, if Eve measures Alice's state in the `correct' basis (implying Eve and Alice share the same bit, $b_E = b_A$ in that slot), her faked state has two possibilities for Bob: 
\begin{enumerate}
	\item Record a detection with $b_B = b_E$ if Bob's chosen basis coincides with Eve's basis. 
	\item No detection event if Bob chooses the incompatible basis.
\end{enumerate}
Figure~\ref{ffsa}b displays these two measurement scenarios (with detectors D0 and D1 assumed to be gated SPADs) due to faked states prepared in the same basis as that used by Alice. 

However, if Eve measures Alice's state in the incorrect basis, the situation is reversed in the sense that a detection of Eve's state in Bob would happen only when Bob's and Alice's bases do not coincide. Since such measurement outcomes are sifted (see subsection~\ref{QKD:rawkey}), the only slots kept for further classical processing would be where Bob registered a detection outcome, and the bases of Alice, Bob and Eve were the same. With the above arguments, it can also be observed that the bits $b_E = b_A = b_B$ in all such slots. In other words, Eve's raw key matches that of Alice and Bob. Thereafter, she can simply listen to error correction and privacy amplification performed by Alice and Bob and apply the same operations on her raw key.

To understand what detector imperfections may enable such an attack, note that SPADs are sensitive to single photons only when they are operated in the Geiger mode. Conversely, in the linear mode, the SPADs are not photon-sensitive; they actually act as normal photodiodes that output currents proportional to the intensity of the input optical pulses. A faked-state attack may then be implemented using pulses of appropriate intensities if Eve can somehow access Bob's detectors in linear mode. 

Before we discuss the methods to do so, note that the intensity levels must be carefully calibrated to control the detection events: different Bob-Eve basis choice slots should \emph{never} result in clicks, identical basis choice slots should \emph{always} give clicks. This is illustrated by Fig.~\ref{ffsa}b. With perfect control, the attack can satisfy the conditions mentioned in subsection~\ref{qh:Conds} easily. In fact, it can even be \emph{completely traceless}~\cite{Lydersen2010} because in principle it does not contribute to the incurred QBER. Secondly, Bob cannot distinguish (at least, not in a simple way) between the genuine quantum signals and faked states because the detection statistics can be preserved. 
%############################################################%
\subsubsection{Blinding loophole}
The notable feature of the first few faked-state attacks~\cite{Lydersen2010,Lydersen2010a,Wiechers2011,Gerhardt2011} was on the ability of Eve to \emph{remotely} create access to Bob's SPADs in the linear mode. In Refs.~\cite{Lydersen2010,Gerhardt2011} for instance, Eve sends in CW light from the quantum channel to Bob, thereby eliciting a current through the SPADs. The impedance of the high-voltage supply (connected to the SPAD for biasing) experiences a voltage drop due to this current. This effectively leads to a lowering of the reverse bias across the SPAD: at a specific CW power, the reverse-bias voltage drops below the breakdown voltage; refer Fig.~\ref{APD_I_V_Diagram}. This implies that the SPAD is out of Geiger mode; or \emph{blind to single photons}. The blinding loophole is arguably the most famous example of quantum hacking; perhaps because it has also been demonstrated on a variety of single photon detectors, including those based on superconducting nanowires~\cite{Makarov2011}, gated SPADs~\cite{Lydersen2010}, and both actively- and passively-quenched SPADs~\cite{Gerhardt2011,Makarov2009,Sauge2011}.
%############################################################%
\subsubsection{After-gate attack}
\begin{figure}%[!htb]
\centering
\includegraphics[width=0.4\textwidth]{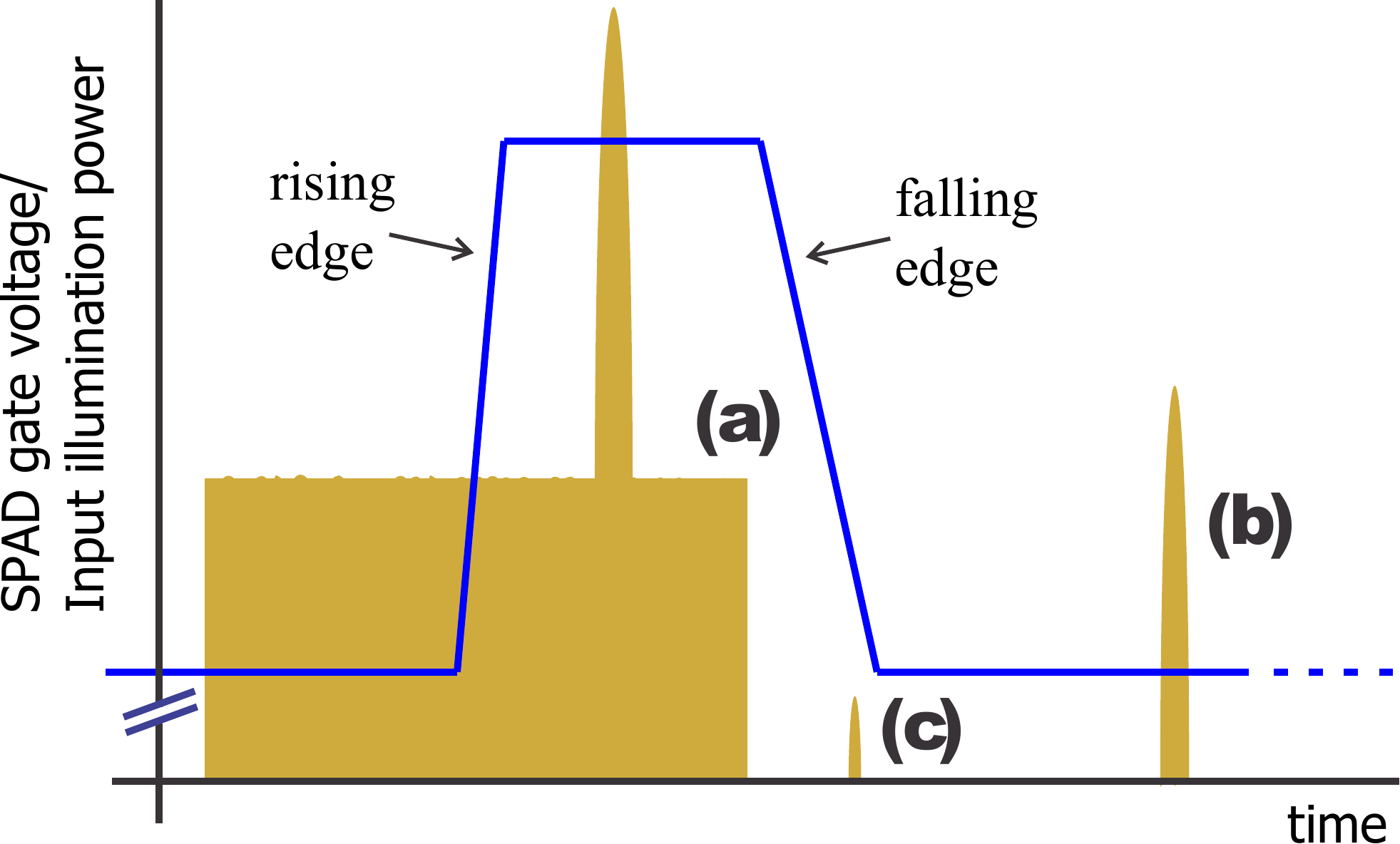}
\caption{(Color online) Faked-state pulses for exploiting the blinding and superlinearity loopholes and used in the after-gate attack on gated SPADs. The blue line show the gate voltage as a function of time. The filled objects represent three different methods for launching the faked-state attack. (a) A bright faked-state pulse containing a few million photons on average arrives inside the gate on top of CW light that has blinded the SPAD. (b) Eve's pulse in the after-gate attack is roughly the same intensity as in (a) but arrives after the falling edge of the gate, i.e. when the SPAD is no longer in Geiger mode. (c) In the superlinearity loophole, the faked-state pulse is relatively dim ($10-100$ photons per pulse) and arrives at the falling edge of the gate.  
\label{ffsacomp}}
\end{figure}

In case of gated SPADs, the linear mode can be remotely accessed in a much simpler way, i.e. without blinding. Typically, the duty cycle of the electronic gating signal is very small, e.g., it is $< 2\%$ for the Clavis2 system~\cite{clavis2DS}. Eve can therefore attack by sending the bright faked-state pulses to impinge on the detector \emph{after the gate}, i.e. when the system has withdrawn the gate (but may still validate the registration of a click due to the impinging bright pulse). Figure~\ref{ffsacomp}b conveys the idea, especially relative to the case of blinding, where the faked-state pulse arrives well inside the gate. 
Indeed, in Ref.~\cite{Wiechers2011}, detector control through bright `after-gate' faked states was demonstrated. However, the bright pulses in this case led to a severe increase in the dark noise of the detectors --- resulting in a rapid rise of the QBER. Nonetheless, simulations showed that this problem could be overcome in at least some operating regimes of the QKD system. The channel transmittance $T$ could also be preserved. 
%############################################################%
\subsubsection{Superlinearity loophole}
\label{superlin}
The optical intensities required to carry out the above attacks are tremendously higher than what Bob normally expects. Simple countermeasures called \emph{watchdog monitors}, which we shall discuss in more detail in section~\ref{cmrs}, would catch such attacks. However, if Eve could somehow use relatively dim faked states, such countermeasures might fail. Of course, for the attack to still work as explained in the beginning of this section, the detector response needs to be somewhat amplified so that even pulses that are $4-5$ orders of magnitude dimmer than their blinding or after-gate counterparts can provide a reasonable level of detector control.

Such an amplification of the response, coined \emph{superlinearity}
~\footnote{The expected behavior scales as $1-\textrm{exp}[-\mu \cdot \eta(t)]$ with the mean photon number $\mu$ of a coherent state used for faking the detection outcomes; $\eta(t)$ is the single photon detection efficiency as a function of time. Since $\eta(t) << 1$ at the falling edge of the gate, $1-\textrm{exp}[-\mu \cdot \eta(t)] \approx \mu \cdot \eta(t)$. In other words, the expected behaviour must be nearly linear. The measured behaviour of the detection probabilities is however always above this linear curve, therefore the term superlinear.}
in Ref.~\cite{LydersenJain2011}, is indeed possible if the dim pulse is sent during the transition from Geiger mode to the linear mode. An investigation of the gated SPADs of Clavis2 in the same work revealed a high degree of superlinearity: a faked-state pulse containing $10-100$ photons and arriving on the falling edge of the gate, as depicted in Fig.~\ref{ffsacomp}c, could be detected with a probability much higher than the theoretically-modelled value. The detector control however is not perfect: while different Bob-Eve basis choice slots sometimes result in clicks (that are erroneous in half the cases), identical basis choice slots do not always yield clicks. The former leads to an increment in the QBER while the latter brings down the overall channel transmittance. Nonetheless, by tuning the parameters of the faked states, QBER and $T$ could be controlled in order to satisfy the conditions given in subsection~\ref{qh:Conds}. 
%############################################################%
\subsection{Laser damage attack}
Instead of coming down the power scale, the eavesdropper can go one step further in the other direction, i.e. even beyond detector blinding. The intention is to inflict optical damage on a certain component in the QKD system and permanently change its characteristics. If the new characteristics then assist the eavesdropper in an attack without Alice or Bob getting any hints of it, the security of that practical QKD system would be in serious trouble. 

In a recently-demonstrated work~\cite{Bugge2014}, the properties and functionality of single photon avalanche diodes (SPADs) were studied after exposing them to laser light with power in the few-watts regime, i.e. $3-4$ orders of magnitude higher than the power levels used in blinding~\cite{Makarov2011,Lydersen2010,Gerhardt2011,Makarov2009,Sauge2011}. As the input power on the SPAD was increased, a large variety of effects --- ranging from a permanently-induced lowering of the efficiency and dark noise level to catastrophic damages to its physical structure --- were noticed. Several loopholes due to these effects were proposed and analyzed. 

One of the attacks is based on the ability of the eavesdropper to \emph{remotely control} the detector efficiency and dark noise levels, which conflicts with the assumptions in most security proofs~\cite{Lutkenhaus2000,Niederberger2005,Branciard2005,Scarani2009a} since these parameters are implicitly assumed to be outside Eve's control. Some proofs in fact suggest to `calibrate' the dark noise beforehand and subtract its effect from the QBER to effectively increase the maximum channel length; see subsection~\ref{realSrcDets}. If Eve could remotely lower the dark noise of Bob's detectors, the resultant decrease in the QBER can be leveraged to attack the system (in some other way, e.g. using faked states).

At very high power levels, the SPAD was found to be damaged to the extent that its interconnects had melted, leaving the component in an open circuit essentially. This suggests that an SPAD based watchdog monitor, deployed at the entrance of Alice/Bob could be rendered useless by shooting enormous levels of optical power into the physical QKD system from the quantum channel. A heightened vulnerability to faked-state attacks was also observed due to the blinding effect, which was remarkably permanent in this case. 

Finally, in some cases, the detection efficiency of the SPAD was also observed to have substantially decreased. Given a quadruple detector assembly, such as the one in Fig.~\ref{BB84BobSetups}b, Eve can target one of the detectors (by choosing the input polarisation) and reduce its efficiency w.r.t. the detector meant to measure the orthogonal polarisation. This opens up the detector efficiency mismatch loophole, which we describe below.
%############################################################%
\subsection{Detection efficiency mismatch loophole}\label{Lnaip:dem}
\begin{figure}%[!htb]
\centering
\includegraphics[width=0.5\textwidth]{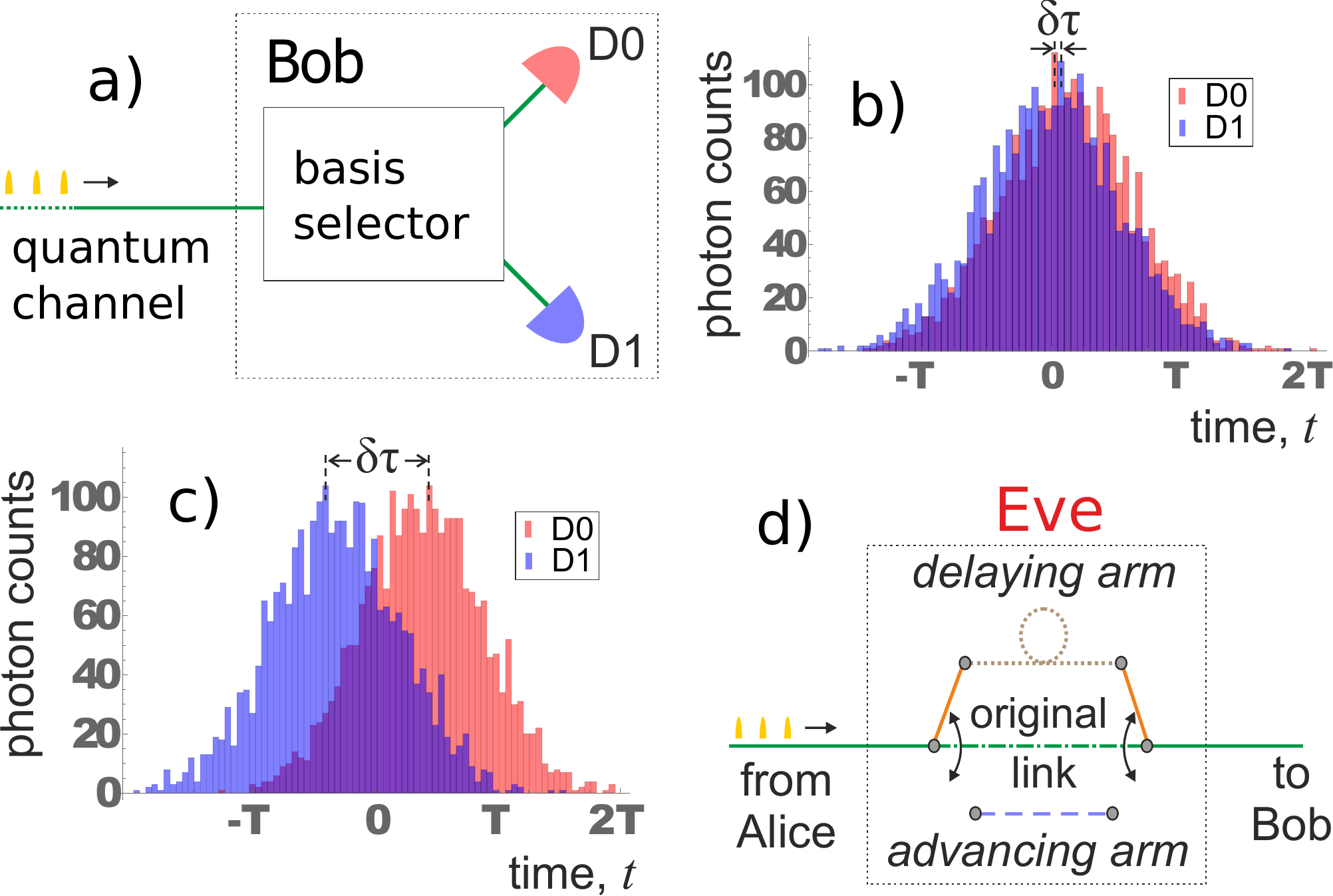}
\caption{(Color online) Detection efficiency mismatch and time-shift attack. a) Bob measures the periodic train of photons from Alice by selecting bases and measuring the corresponding outcome in D0 or D1 for each photon. b) A simulated histogram of the instants in time when photodetection events are recorded by D0[D1] in red[blue]. While not ideal, the situation is still secure because to Eve, no distinguishing information is available. c) The separation in time between the red and blue histograms results in a mismatch that Eve can exploit in different ways. d) Schematic of the equipment to attack the insecure system shown in c) by delaying/advancing the arrival time of Alice's photons in Bob.
\label{fdemtsa}}
\end{figure}

The origin of this loophole lies in the relationship of Bob's measurement results with his detectors. Theoretically, and as explained in subsection~\ref{QKD:rawkey}, only the relative choice of the bases of Alice and Bob must determine the measurement outcome. Assuming the detector assembly in Fig.~\ref{fdemtsa}a, this would require D0 and D1 to be indistinguishable. This is hard to obtain in practice since the physical properties of two detectors cannot be the same. Limitations in the manufacturing process along with variations in environmental conditions can make the detectors discernible~\footnote{To understand the concept of discernability in a simple way, one should be able to, for instance, simply exchange D0 and D1 without hindering or altering the operation of the QKD system. Of course, the key obtained by Bob would then be uniformly opposite to that of Alice but a simple NOT operation would suffice to reconcile their keys.}. For example, lengths of the fibers connecting D0 and D1 can vary~\cite{Makarov2006} due to different temperatures, jitter produced by an electronic circuit fluctuates over time, sensitivities at the same input wavelength can differ~\cite{Fung2009} due to intrinsic variations in the material composition. 

If the discerning physical property also aids control over --- the clicking of --- the mismatched detectors, a QKD system that employs that pair of detectors will carry a risk of an attack due to the detection efficiency mismatch (DEM) loophole~\cite{Makarov2006}. Figures~\ref{fdemtsa}a--\ref{fdemtsa}c explain the idea of the loophole, with the discerning physical property being time.
In an ideal world, a periodic train of incoming photons would be detected and counted at precisely the same time (say $t=0$, measured relative to a reference, such as the system clock) in both D0 and D1. In practice, however, there is a spread as illustrated by the red and blue histograms of the arrival times measured by D0 and D1 in Fig.~\ref{fdemtsa}b, with $T$ denoting the full width at half maximum (FWHM) value. Nonetheless, the large overlap between the two histograms signifies that the detectors are not easily discernible. 

The outer envelopes of the histograms correspond to the detection efficiencies $\eta_0(t)$ and $\eta_1(t)$ of D0 and D1, respectively, as a function of time. The efficiency mismatch that allows distinguishing the detectors strongly depends on the temporal separation $\delta \tau$ between the efficiency curves. This can be easily visualized by comparing Figs.~\ref{fdemtsa}c and \ref{fdemtsa}b. 
%The amount of efficiency mismatch can be quantified by the logarithm of their ratio, $\epsilon(t) = \log_{10}[\eta_0(t) / \eta_1(t)]$. It may be observed that the function $\epsilon(t)$ strongly depends on the temporal separation $\delta \tau$ between the efficiency curves. In the situation shown in Fig.~\ref{fdemtsa}c, one may visualize the absolute value of the function $|\epsilon(t)|$ reaching a maximum around time $t_{\pm}=\pm \delta \tau/2$. 

Two attacks for exploiting the DEM loophole are known. In the original proposal~\cite{Makarov2006}, an attack using faked states~\cite{Makarov2005a} was considered and its performance was theoretically analyzed. A simpler alternative which we discuss below is the time-shift attack~\cite{Zhao2008}. Notably, this was the first known quantum hacking attempt on a commercial QKD system. 
%############################################################%
\subsection{Time-shift attack\label{Lnaip:tsa}}
Suppose Eve is aware that the QKD implementation suffers from detection efficiency mismatch as illustrated in Fig.~\ref{fdemtsa}c. To attack the system, she can simply delay or advance the arrival of the quantum signal pulses in Bob's subsystem randomly. 
%Some photons arrive at $t \approx t_{+}$ while others arrive at $t \approx t_{-}$ as a result. 
Figure~\ref{fdemtsa}d shows an apparatus constructed using optical switches and fibre patchcords that can perform the above actions, once installed in the quantum channel. 

Using a sufficiently long[short] fibre patch cord to act as the delaying[advancing] arm, Eve can make sure that the probability for Bob to detect a photon in D1[D0] is negligible \emph{even if his basis choice relative to Alice's should have resulted in that outcome}. Due to this, Bob's measurement outcomes are highly dependent and correlated with Eve's actions. The keys obtained by Bob and Alice would be quite similar to that of Eve (\emph{delaying} $\equiv$ bit $0$ and \emph{advancing} $\equiv$ bit $1$). 

Nonetheless, one can realize that the time-shifting action results in fewer detection events than normally expected. This impacts the estimation of the channel transmittance and implies that the second condition listed in subsection~\ref{qh:Conds} may be hard to satisfy. Also, the incurred QBER would elevate as the `photonic' detections are lowered with respect to the intrinsic `dark noise' of the detectors; see subsection~\ref{realSrcDets}.

More practically, the assumption that Eve knows the exact nature of detection efficiency mismatch may not hold true. In the actual demonstration~\cite{Zhao2008}, the authors (who obviously had access to the QKD system) performed the time shifts only in the instances where the efficiency mismatch  favoured their attack. But in practice, Eve does not have access to the actual system and there is no information that allows her to control or predict the temporal separation $\delta \tau$. Even worse, $\delta \tau$ is a stochastic quantity fluctuating around zero. Figure~\ref{fcalibhack} illustrates this scenario (only the green disks) for Clavis2; we shall explain this in more detail below. Its impact on the efficacy of a realistic time-shift attack should however be clear: instances in which $|\delta \tau|$ is large (as marked by the green disks with red borders) occur in $<5\%$ of all instances. 
%############################################################%
\subsection{Calibration loophole\label{Lnaip:calib}}
\begin{figure}%[!htb]
\centering
\includegraphics[width=0.45\textwidth]{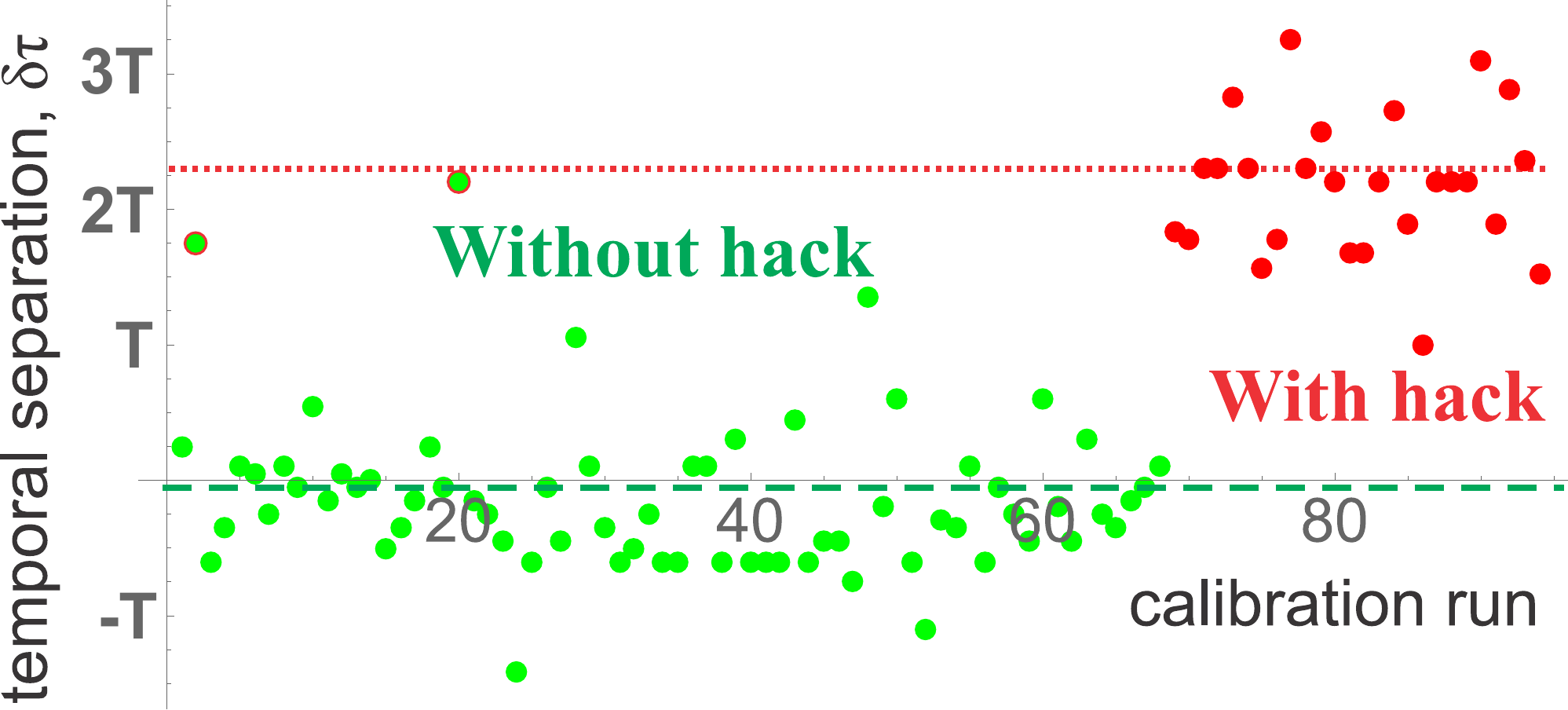}
\caption{(Color online) Shift in the temporal separation between the detectors of Clavis2 due to Eve's manipulation. Experimentally measured values of the temporal separation obtained after calibration runs operated without (in green) and with (in red) Eve's hack; $T$ is the FWHM value of the distributions shown in Figs.~\ref{fdemtsa}b and \ref{fdemtsa}c. The induced detection efficiency mismatch (DEM) after the hack is sufficiently high to increase the efficacy of a time-shift attack~\cite{Zhao2008}, or even launch a successful faked-state attack~\cite{Jain2011}.
\label{fcalibhack}}
\end{figure}

To use the time-shift attack successfully, Eve needs the QKD system to somehow exhibit a large temporal separation. Additionally, the corresponding detection efficiency mismatch must remain deterministic and ideally, the variance of $\delta \tau$ must be as low as possible. The right hand side of Fig.~\ref{fcalibhack} (i.e. only the red dots) depicts a scenario in which at least the first condition is satisfied. 
These points were experimentally observed on the Clavis2 QKD system~\cite{clavis2DS} by repetitively running by the detector calibration routine in a manner that remotely exploited a loophole in the routine firmware and led to the \emph{induction} of this temporal separation~\cite{Jain2011}. 

It can be compared with the situation when the calibration routine was run normally (green dots in Fig.~\ref{fcalibhack}). Note that while the corresponding average value of $\delta \tau$, shown by the dashed-green line, is indeed quite close to zero, a small fraction of calibration runs did exhibit high detection efficiency mismatch. These instances are depicted by green dots with red borders.

The calibration routine is run by Clavis2 intermittently between the key exchanges to reduce any existent DEM, which can occur over time due to reasons explained in subsection~\ref{Lnaip:dem}. The basic steps are: Alice sends a train of classical light pulses (instead of quantum signals) to Bob over the quantum channel. Bob periodically scans in time the arrival of Alice's pulses by changing the gate-activation times of both detectors independently. As and when the activation instant coincides with the impinging of the pulse on say D0, the number of photon detection events recorded in D0 reaches maximum. A similar maximum in D1 is recorded; the final situation \emph{ideally} looks like the one presented in Fig.~\ref{fdemtsa}b. The corresponding value of $\delta \tau \approx 0.0$ on average, as illustrated by the green-dashed line in Fig.~\ref{fcalibhack}.

The weakness that allowed the exploitation of this loophole arose due to the \emph{fixed settings of bases} by Alice and Bob during the calibration phase~\cite{Jain2011}. In the normal case, photons from any part of Alice's optical pulse can yield a click in Bob's detectors. However, Eve can manipulate these classical pulses (on the quantum channel) in such a way that the first temporal half yields clicks only in D1, while the second half yields clicks only in D0. This results in a situation similar to that shown in Fig.~\ref{fdemtsa}c. Experimentally, an average separation $\delta \tau > 2T$ is achieved, as illustrated by the dotted-red line in Fig.~\ref{fcalibhack}. This large efficiency mismatch induced between Bob's detectors can be exploited by a faked-states attack and the conditions of subsection~\ref{qh:Conds} can be satisfied for a large range of the channel transmittance values. 
%############################################################%
\subsection{Wavelength-dependent beam splitter attack\label{Lnaip:wvdep}}
Imperfections in components other than detectors can also result in vulnerabilities. A recent example~\cite{Huang2013,Li2011} is due to the wavelength dependence of a beam splitter, which is one of the most frequently-used components in quantum-optical experiments. A symmetric beam splitter employed at the telecom wavelength of $\lambda \approx 1550\,$nm would have roughly the same reflectivity and transmittivity values, i.e. $R(\lambda) \approx T(\lambda) \approx 0.50$ (the absorption losses are usually insignificant). A photon passing through this beam splitter would hence be detected in the transmissive or the reflective output with equal probability. 

The response of the same beam splitter at other wavelengths can however be quite different. Let us assume there are two wavelengths $\lambda_0 \neq \lambda_1$ at which the beam splitter is highly transmissive and highly reflective, respectively, i.e. $T(\lambda_0) \lesssim 1.0$ and $R(\lambda_1) \lesssim 1.0$. An attacker can substitute the quantum signals at $\lambda$ with light at $\lambda_0$[$\lambda_1]$ to address only the transmissive[reflective] port in a \emph{deterministic} fashion. More specifically, an intercept and resend attack that exploits the wavelength dependence is possible. 
\begin{figure}%[!htb]
\centering
\includegraphics[width=0.5\textwidth]{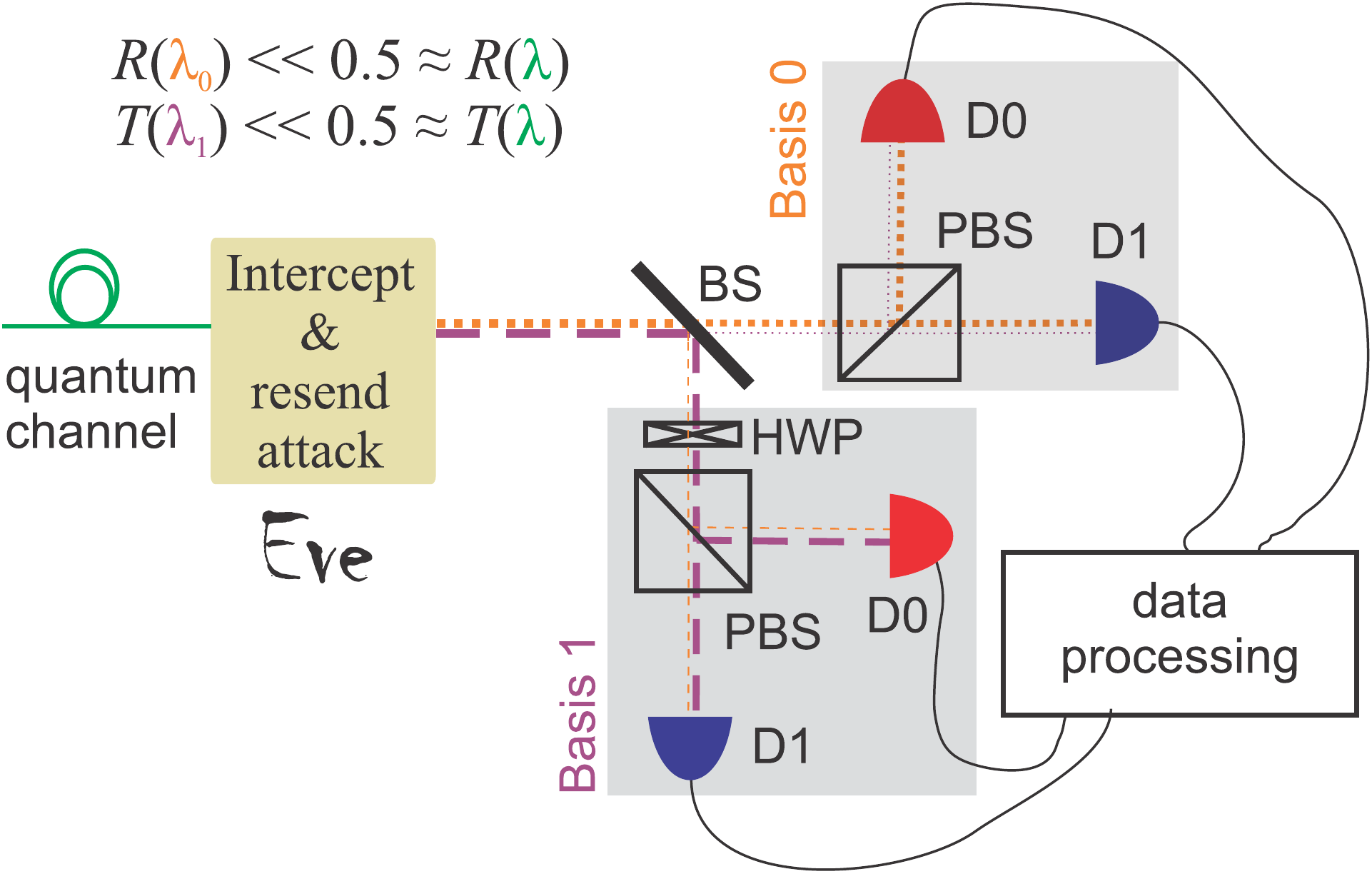}
\caption{(Color online) Intercept and resend strategy to exploit the wavelength dependence of a beam splitter. Eve intercepts and measures the photon from Alice in Basis $j$ ($j = 0$ or $1$). The passive beam splitter (BS) in Bob has a very different splitting ratio at $\lambda_j$ with the values $R(\lambda_0) = 0.3 \times 10^{-2}$ and $R(\lambda_1) = 0.986$ experimentally observed in Ref.~\cite{Li2011} for $\lambda_0 = 1290\,$nm and $\lambda_1 = 1470\,$nm. Eve's resent state (not necessarily quantum in nature) to Bob is prepared at $\lambda_j$ which means that the chances of Bob measuring it in the Basis $i$ are quite low[high] if $i \neq j$[$i = j$]. Also, Eve can easily select the detector to be clicked (D0 or D1) by choosing an appropriate polarisation for her resent state. 
\label{fWvlDep}}
\end{figure}
Figure~\ref{fWvlDep} shows the scheme of such an attack demonstrated with $\lambda_0 = 1290\,$nm and $\lambda_1 = 1470\,$nm~\cite{Li2011}.

The attack reportedly did not increase the QBER measured by Bob in any significant manner. Although the detection efficiency of Bob's SPADs also varied at $\lambda_0$ and $\lambda_1$, any differences in the detection rates (with respect to the values expected at the normal wavelength $\lambda$) could be compensated by employing appropriately-bright coherent states in the resending stage.
%############################################################%
\subsection{Trojan-horse attacks\label{Lnaip:trojha}}
Trojan-horse attacks are amongst the most well-known class of attacks on practical QKD implementations. This term has been borrowed from classical cryptography although in the context of practical QKD, it is a misnomer since most attacks demonstrated in this category \emph{do not} involve Alice or Bob accepting seemingly-benign objects from Eve. Coined in Ref.~\cite{Gisin2006}, the term has become well established over the years even though another term `large pulse attack' for the same concept had been proposed~\cite{Bethune2000a,Vakhitov2001}. Nonetheless, to maintain consistency, we shall not buck the trend here.%, e.g., optical pulses designed by Eve to launch an attack in this subsection will be referred to as Trojan-horse pulses. 

In principle, the basic idea facilitating these attacks is that light is reflected/scattered back as it propagates through optical components. For instance, the change of refractive index across an interface between two components results in a Fresnel reflection. In case the light encounters mirror-like surfaces, a specular reflection is obtained. If the back-reflected photons travel through a basis selector, such as those shown in Figs.~\ref{fBB84AliceBob}b and \ref{BB84BobSetups}a, they can capture the state of the polarizer or the modulator. If they also eventually trickle out of the QKD system unnoticed onto the quantum channel, Eve can intercept them and perform appropriate measurements to discriminate between different bases settings. 

Eve can actively create grounds for such an attack by sending bright pulses into the equipment of Alice/Bob from the quantum channel and scanning through the different reflections to obtain the relevant information. The works in Refs~\cite{Vakhitov2001,Gisin2006} explored the basic ideas and applicability conditions for implementing such type of attacks. They also proposed experimental tools, e.g. optical time domain reflectometers, to quantify the reflections in time and amplitude of a fibre-optical system. This facilitated building of the (static) reflection maps of sample QKD systems which could be crucial to decide Eve's photon budget. Recently, proof-of-principle attacks were carried out in real-time on the QKD research platforms of ID~Quantique and SeQureNet~\cite{Jain2014,Khan2015}. The basic message of these attacks is that Eve can probe the settings of the basis selector on-the-fly and just with a handful of the back-reflected photons. 

Eve's attack may be caught using a so-called watchdog monitor, which we shall describe in some detail in the next section. Alternatively, an optical isolator can thwart the attack by simply preventing light from Eve to even enter the QKD system. However, apart from the fact that such countermeasures cannot always be simply adopted for a given QKD system, they may anyway prove vulnerable if Eve resorts to another wavelength for the attack~\cite{Jain2014,Jain2015}. 
%The location of components, especially the basis selector, inside Alice/Bob can hence be quite critical in the context of Trojan-horse attacks. For instance, as per figure 2 in Ref.~\cite{Silva2013}, an automatic polarisation controller (APC), most likely acting as the basis selector, is placed at the entrance of Alice. This can be hazardous for the security because any reflection occuring inside the rest of the subsystem would pass through the APC and reveal its chosen state. 
%############################################################%
\section{Countermeasures}\label{cmrs}
In the previous pages, we have demonstrated ample cases of how a quantum hacker exploits deviations between the theoretical model and the practical implementation to attack the QKD system. We now discuss approaches and mechanisms --- technically referred to as countermeasures --- to prevent or catch such attacks from breaking the security, thereby restoring the security of the practical QKD system. 
%############################################################%
\subsection{Basic approaches}
Against side channels in particular, two approaches are possible: eliminate the leakage of information or make it useless for the adversary. The first is usually quite difficult, a better practice is to try binding the leakage and performing higher privacy amplification. Alternatively, the correlations between the side channel information and the actual secret encoding/decoding data can be reduced to the minimum possible level~\cite{Nauerth2009}. 
%############################################################%
\subsubsection{Addition or modification in hardware/software}
\begin{figure}%[!htb]
\centering
\includegraphics[width=0.5\textwidth]{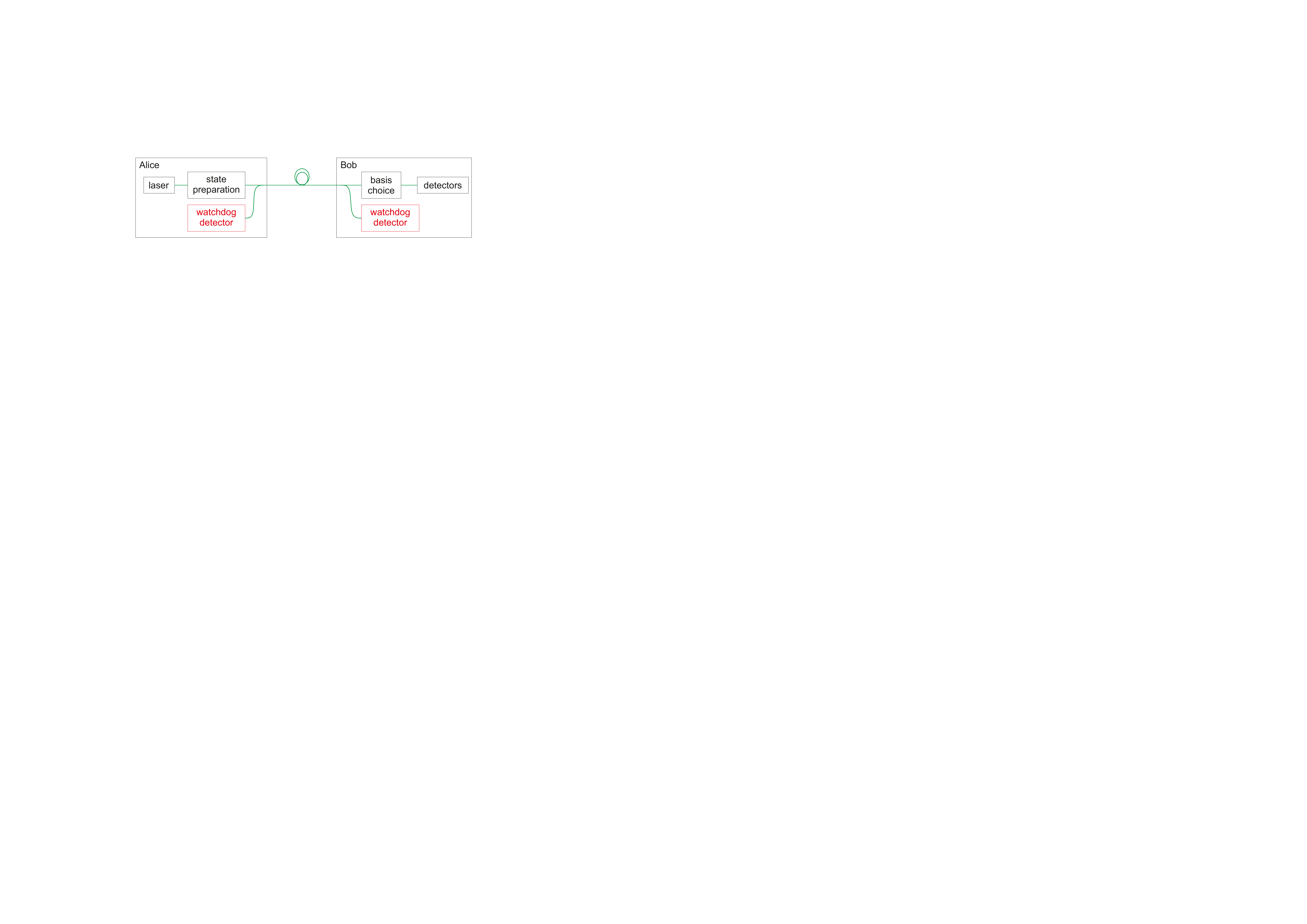}
\caption{(Color online) Basic QKD implementation with watchdog detectors to guard against Eve's radiation. The detectors monitor the incoming light to find out if Eve is carrying out an attack. These monitoring devices may be connected to the rest of the QKD system via a coupler or a switch. Ideally, a watchdog detector should be able to discern legitimate quantum signals from Eve's radiation to prevent false alarms. \label{WatchdogDetectors}}
\end{figure}

For a significant number of attacks on QKD systems discussed above, Eve employs bright light pulses. In the cases of blinding and after-gate attacks~\cite{Makarov2011,Lydersen2010,Gerhardt2011,Wiechers2011}, she sends the bright light towards Bob, targeting the imperfections of the detection system. In a Trojan-horse attack, Alice's device could also be subject to the bright illumination. For one-way QKD systems, an optical isolator at Alice's entrance can be used to block the light from Eve. A more universal method consists in placing an additional detector at the entrance of Alice's and Bob's device. This so-called watchdog detector ideally monitors a portion of all incoming radiation; see Fig.~\ref{WatchdogDetectors}. 
When the incoming light intensity is above a certain threshold, the QKD operation is interrupted and an alarm is raised. 

In practice, a compromise is required between the fraction of light being monitored by the watchdog detector and the losses induced by the tap. Randomly switching between using the incoming light for QKD operation and for monitoring offers an alternative~\cite{Jain2014}. Another method would be to use the quantum light detectors themselves to monitor unwanted incoming light. As reported in Ref.~\cite{Silva2012}, traces of several attacks can be found by carefully analysing the detection statistics from the quantum detectors. Instead of additional hardware, the countermeasure can therefore be implemented in a software algorithm. A software countermeasure can also repel the calibration loophole described in section~\ref{Lnaip:calib}. The software patch consists in implementing a random basis choice in Bob during the calibration routine as well~\cite{Jain2011}.

Changes in hardware are usually considered the last resort. Apart from the obvious investment of time, effort, and money in installing and testing the new hardware, such countermeasures can also prove to be the source of new risks. Alternatively, they may even prove insufficient in protecting the QKD system. For example, it is vital that both isolators and watchdog detectors function in the \emph{desired} manner at ideally all the wavelengths that Eve could use to attack Alice's or Bob's devices~\cite{Jain2015}. However, due to intrinsic limitations, as explained further in subsection~\ref{cmrs:intm} below, this is quite difficult to guarantee. In other words, such countermeasures would provide Alice and Bob with a false sense of security.
%############################################################%
\subsubsection{Incorporating imperfections in security proof}
In some cases, the imperfections of the QKD system can be theoretically quantified. An example is the loophole arising due to the different spectral distributions of the individual laser diodes used to prepare the quantum state alphabet~\cite{Nauerth2009}. In this case, information leaked to Eve can be erased by applying more privacy amplification. This effectively shrinks the key but with the assurance that Eve's partial information is destroyed. Measures like these are highly beneficial for practical QKD since by directly considering the loophole in the security proof, they reduce the theory-practice deviation. Recently, quantitative bounds for a system exposed to Trojan-horse attacks were reported, and a passive architecture to counteract the attack was proposed~\cite{Lucamarini2015}. The key element in the evaluation -- based on the specifications of realistic optical components used in QKD -- was to treat the Trojan-horse attack as a side channel.
%############################################################%
\subsection{Conceiving novel QKD schemes}
Instead of patching existent QKD protocols, it is also possible to devise schemes that are immune to one or more classes of hacking attacks. Device independent QKD (DI-QKD) is based on the violation of a Bell inequality in an entanglement-based QKD scheme~\cite{Acin2007}. Such a violation can occur only when Alice and Bob share quantum correlations. Proving a Bell inequality violation relieves Alice and Bob from the need to exactly characterise the imperfections of their sources and detectors in the context of possible attacks. However, DI-QKD currently suffers from major practical drawbacks. The tolerable losses are so low that a nearly loss-less channel (in combination with highly-efficient detectors) is required.

The extreme requirements of DI-QKD can be relaxed when employing the concept of measurement device independent QKD (MDI-QKD)~\cite{Lo2012,Braunstein2012}. Here, the sources of Alice and Bob are assumed to be outside the reach of Eve, whereas the detectors need not be trusted and can even be controlled by Eve. (More generally, the idea of public-private spaces of Alice and Bob and untrusted relays in the middle to perform quantum measurements -- on the public systems of Alice and Bob -- can be used for designing QKD schemes that are free of all side channels~\cite{Braunstein2012}.) This is a remarkable development for QKD since the majority of attacks so far exploited the imperfections in the detection apparatus. In the MDI-QKD scheme, Alice and Bob send quantum states to an untrusted relay called Charles who performs a Bell state measurement. The relay is potentially controlled by Eve, its trustworthiness is checked publicly by analysing the outcomes of the Bell state measurements. This scheme tolerates losses in the same order as standard QKD protocols and is therefore practically feasible~\cite{Yang2013}. 

Another idea that can address numerous vulnerabilities in the photon detection apparatus of realistic QKD implementations is to use optical sum frequency generation, or upconversion, to protect Bob~\cite{Jain2016}. The process of upconversion employs a nonlinear interaction between the quantum signal (from Alice) and a bright pump (inside Bob), and the characteristics of the pump may be manipulated/monitored by Bob to prevent/detect a wide range of quantum hacking attacks launched by Eve. In that sense, such an upconversion protected QKD receiver is similar to MDI-QKD, though unlike the latter, it lacks a rigorous security proof. 

Finally, continuous-variable (CV) QKD schemes employ homodyne detection of light using photodetectors that operate in the linear regime. Therefore, the faked-states attacks described in subsection~\ref{Lnaip:fsas} are not straightforwardly implementable on CV-QKD systems. However, in a majority of CV-QKD setups, an additional reference beam travels through Eve's domain. Care has to be taken that Eve's potential manipulations of this reference are monitored~\cite{Haeseler2008,Wittmann2010,Huang2014}.
%############################################################%
\subsection{Word of caution against intuitive measures \label{cmrs:intm}}
%############################################################%
\subsubsection{Isolators against Trojan-horse attacks}
As mentioned above, optical isolato is one of the best-known measures to shield practical QKD systems against Trojan-horse attacks. Such a device lets light pass when it comes from the forward direction, but blocks it when coming from the reverse direction. However, it was recently discovered that the light blockage can be guaranteed only in a narrow region around the design wavelength~\cite{Jain2015}. Eve can therefore just choose another wavelength where the extinction level of the isolator (in double pass) is insufficient. This phenomena occurs since typical isolators are based on the wavelength-dependent Faraday effect. A solution to this vulnerability is to use the optical isolator in conjunction with a wavelength filter~\cite{Jain2015}. 
%############################################################%
\subsubsection{Watchdogs (superlinearity loophole)}
The superlinearity loophole as described in subsection~\ref{superlin} can be exploited by dim states of light injected by Eve into Bob's device. Therefore this attack is difficult to catch with standard powermeter based watchdog detectors monitoring the incoming light power. The watchdog detector itself would have to be sensitive on the few photon level, exposing it to loopholes discussed in section~\ref{Lnaip:fsas}. A remedy to this loophole consists in applying gates with a varying activation time or width~\cite{Legre2010}. Another possibility consists in the modification of the mapping between detectors and bit value~\cite{Lydersen2011a}. This remapping is done both in software and in the optical system by applying a phase shift of either $0$ or $\pi$. This so-called bit-mapped gating leads to a QBER of 50\% for detections outside the middle of the gate.
%############################################################%

\section{Take-home message}%Conclusion
As it celebrates its 30th anniversary, quantum key distribution (QKD) has evolved from being just `a beautiful idea' to the first applied quantum technology~\cite{Bennett1984,Gisin2002,Scarani2009a,Lo2014}. However, as and when a technology matures, bugs and loopholes are imminent. Quantum hacking addresses the deviations between the theoretical and practical worlds of QKD by discovering and closing loopholes and thus should be considered as a natural as well as essential ingredient in the evolution. Let us emphasize that the role of quantum hacking is not to derail its progress but to prevent security problems in future if and when QKD is deployed. That fortunately has also been the case so far: the rise of quantum hacking has also gone hand in hand with miscellaneous landmarks in the last decade. To name a few:
\begin{itemize}
	\item Maximum possible distance between Alice and Bob has surpassed lengths well over $100\,$km~\cite{Ursin2007,Schmitt-Manderbach2007,Takesue2007,yuan2009,Jouguet2012} and satellite quantum communication is in development at several places in the world. (Depending on the application scenario, different orbital distances have been considered~\cite{Jennewein2013,Dequal2015,Elser2015}).
	\item QKD links between rapidly-moving or floating platforms have been demonstrated~\cite{Nauerth2013,Wang2013}.
	\item Fiber networks with multiple nodes and new architectures have been established~\cite{Zeilinger2009,Sasaki2011,Frohlich2013}.
	\item Efforts for the standardization of QKD spearheaded by the European Telecommunication Standards Institute (ETSI) are in a fairly mature stage~\cite{ETSI2014}.
\end{itemize}
With advances in quantum computation threatening the security of public key cryptographic systems, it is important that security of technologies such as practical quantum key distribution be scrutinized so that viable alternatives are present in the future.
%############################################################%
\section*{Acknowledgements}
We thank Kevin Guenthner for proofreading the manuscript.
%\selectlanguage{german}
%\theendnotes
\bibliography{library}% Produces the bibliography via BibTeX.
\end{document}